\documentclass[aps,prl,twocolumn,groupedaddress,floatfix,amsmath,amssymb,longbibliography,10pt]{revtex4-1}

\usepackage[utf8]{inputenc}
\usepackage[T1]{fontenc}
\usepackage{txfonts}
\usepackage{tgtermes}
\usepackage[altg]{qtxmath}
\usepackage{braket}
\usepackage{graphicx}
\usepackage{amsmath}
\graphicspath{{figures/}}
\usepackage{color}
\usepackage[english]{babel}
\usepackage{microtype}
\usepackage{url}
\usepackage{pdfpages}

\def\BBraket#1{\left\langle\!\left\langle#1\right\rangle\!\right\rangle}
{\catcode`\|=\active
  \xdef\BBraket{\protect\expandafter\noexpand\csname BBraket \endcsname}
  \expandafter\gdef\csname BBraket \endcsname#1{\begingroup
     \ifx\SavedDoubleVert\relax
       \let\SavedDoubleVert\|\let\|\BraDoubleVert
     \fi
     \mathcode`\|32768\let|\BraVert
     \left\langle\!\left\langle{#1}\right\rangle\!\right\rangle\endgroup}
}

\renewcommand*{\vec}[1]{\boldsymbol{#1}}

\makeatletter
\AtBeginDocument{\let\LS@rot\@undefined}
\makeatother

\begin{document}

\author{E.~Raicher}
\email{erez.raicher@mpi-hd.mpg.de }
\author{Q.~Z. Lv}
\email{qingzheng.lyu@mpi-hd.mpg.de }
\author{C.~H. Keitel}
\author{K.~Z. Hatsagortsyan}
\affiliation{Max-Planck-Institut f\"{u}r Kernphysik, Saupfercheckweg 1,  69117 Heidelberg, Germany }

\title{New class of violation of local constant field approximation in intense crossed laser pulse scenarios}

\date{\today}

\begin{abstract}

It is commonly assumed that in ultrastrong laser fields, when the strong field parameter of the laser field $\xi$ is larger than one, the electron radiation is well described by the local constant field approximation (LCFA). We discuss the failure of this conjecture, considering radiation of an ultrarelativistic electron interacting with strong counterpropagating laser waves. A deviation from LCFA,  in particular in the high-frequency domain, is shown to occur even at $\xi\gg 1$ because of the appearance of an additional small time scale in the trajectory. Moreover, we identify a new class of LCFA violation, when the radiation formation length becomes smaller than the one via LCFA. It is characterized by a broad and smooth spectrum rather than an harmonic structure. A similar phenomenon is also demonstrated in the scenario of an electron colliding with an ultrashort laser pulse. The relevance to laser-plasma kinetic simulations is discussed.

\end{abstract}

\date{\today}


\maketitle

Exploration of novel regimes of laser-matter interaction, including nonlinear QED {\cite{Mourou_2006,Marklund_2006,RMP_2012,Heinzl_2012,Dunne2014,Turcu_2019,Bula_1996,Burke_1997} and radiation reaction \cite{Cole_2018,Poder_2018}} effects,  has been enabled due to the dramatic progress in high-power laser technology \cite{Strickland_2019,Mourou_2019}. The peak power of contemporary lasers has currently attained the petawatt regime \cite{Yoon_2019, Vulcan1}
(corresponding to the  field parameter $\xi = e E_0 /m \omega_0 \sim 100$) and
multipetawatt infrastructures are under construction worldwide
\cite{ELI, XCELS}.  Here, $-e$ and $m$ are the electron charge and
mass, respectively,  $E_0$ and $\omega_0$ are  the laser field amplitude, and the frequency, respectively (units $\hbar = c = 1$ are used throughout).

Strong-field QED processes in laser fields can be treated  fully quantum
mechanically only for limited field configurations,
where the single-particle wave function is available \cite{Volkov_1935,DiPiazza_2014}.
Therefore, the standard Monte Carlo codes, which are employed for theoretical
investigation of QED effects in laser-plasma interaction \cite{Elkina_2011,Ridgers_2014,Green_2015},
treat photon emissions and pair production with the local constant field approximation (LCFA).

The LCFA is commonly believed to be applicable  in strong
laser fields with $\xi\gg1$.  One can formulate the physical  condition for the LCFA applicability as $t_f\ll t_c$, in terms of the formation time $t_f$ of the radiation emission (pair production) and the  characteristic time of the electron trajectory $t_c$ \cite{Baier_b_1994,Ritus_1985}.
As an ultrarelativistic electron emits forwards within $1/\gamma$-cone \cite{Jackson_b_1975}, with
 the Lorentz factor $\gamma$, the radiation can be superimposed and coherently formed during
the time $t_f$ spent by the electron in the $1/\gamma$-cone.  In a circularly polarized
monochromatic plane wave, the trajectory of an electron in the average rest frame is a circle and the corresponding Lorentz factor $\gamma=\xi$ \cite{Sarachik_1970}. When $\xi\gg 1$, the formation length is $\ell_f\sim \rho
\theta \sim \rho/\gamma$, with the radius  of the trajectory $\rho$, and the angle of the coherent emission $\theta \sim 1/\gamma$. The typical length scale of the trajectory in this case
 is $\ell_\mathrm{c}=\rho$.  Therefore, $t_\mathrm{f}/t_\mathrm{c}=\ell_\mathrm{f}/\ell_\mathrm{c} \sim 1/\xi\ll 1$ is
automatically fulfilled at $\xi \gg 1$ in a single laser field, similar to the synchrotron radiation case of  an ultrarelativistic electron at $\gamma\gg 1$.  However, in multiple beam laser configurations
(see, e.g., a promising configuration of the dipole wave \cite{Bulanov_2010_a,Gonoskov_2012,Gonoskov_2017,Magnusson_2019}) several characteristic time scales can appear in the electron trajectory. As a result, the condition
$t_f\ll t_c$ will not be equivalent to $\xi\gg 1$ and therefore, as shown below, violation of LCFA may arise then.

The simplest multiple beam laser configuration is a setup of counterpropagating laser waves (CPW), which exhibits radiative trapping dynamics \cite{Gonoskov_2014,Kirk_2016}, and is favorable for the exploration of nonlinear QED \cite{Kirk_2009,Grismayer_2016,Jirka_2016,Gong_2017,Grismayer_2017,Mueller_2017}. The CPW scheme may arise also in a laser-plasma interaction due to the reflection of the impinging laser wave from the critical density \cite{Brady_2012}.

\begin{figure}[b]
  \begin{center}
  \includegraphics[width=0.32\textwidth]{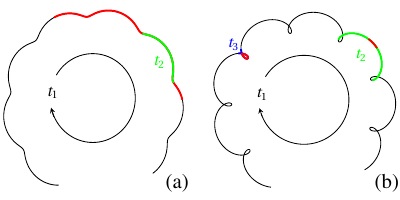}
  \caption{Examples of electron trajectories  in circularly polarized CPW (projection on the plane transverse to the laser propagation):  (a)  $\xi_1\gg 1$  and $\xi_2<1$ ($t_{f2}>t_2$); (b) $\xi_1\gg \xi_2\gg 1$ ($t_{f2}<t_2$, but $t_{f3}\approx t_3 $);  (red) is the radiation formation length, (green) and (blue) are the additional characteristic lengths of the electron trajectory [$t_i $ and $t_{fi}$ ($i=1,2,3$) are the corresponding characteristic and formation times]. While in a single laser wave only the scale $t_1$ for the trajectory is available, in CPW additional small time scales arise. LCFA fails at $t_{fi}\gtrsim t_i$. In (a) $t_{f2}>t_{f2}^L$, while in (b) $t_{f3}<t_{f3}^L$ which represents a new class of LCFA violation.
    }
  \label{fig:illustration}
  \end{center}
\end{figure}

In this Letter we show for CPW of the same frequency, that two additional small characteristic time scales $t_2,\,t_3 $ emerge in the trajectory along with the fundamental one $t_1$, which corresponds to a single laser wave (see Fig.~\ref{fig:illustration}). In case these are comparable to the corresponding formation time scales, $t_{fi} \gtrsim t_i$, $i=1,2,3$, a deviation from LCFA arises.
Since $t_c \equiv |\vec{F}|/|\dot{\vec{F}}|$ where $\vec{F}$ is the Lorentz force, the short time scale $t_3$ corresponds to the parts of the trajectory where the total force is relatively small but changing rapidly.
Furthermore, while in  Fig. 1(a) the formation time is larger than the formation time via LCFA ($t_f^L$),  and hence the LCFA violation is similar to that in a single laser wave, in the case of Fig. 1(b) the electron motion during $t_3$ is extremely abrupt, giving rise to a previously unidentified class of LCFA violation with $t_f<t_f^L$ and an up-shifted typical emitted frequency with respect to LCFA.

One should distinguish between recently found deviations from LCFA in radiation spectra of an electron in a single laser wave with $\xi\gg 1$ \cite{DiPiazza_2018,DiPiazza_2019,Ilderton_2019}, and that
in CPW discussed in this Letter. While here the failure of LCFA is due to violation of the condition $t_f \ll t_c$, as additional small scales $t_c$ arise in the trajectory, in the former, the interference between radiation emerging from different laser cycles creates harmonic peaks visible in the low energy part of the spectrum, in deviation from the LCFA result.

The electron radiation in CPW is explored here in the realm of the Baier-Katkov semiclassical formalism \cite{Katkov_1968,Baier_b_1994}.
This method employs the classical electron trajectory, however, accounts for the quantum recoil due to a photon emission.  We consider two qualitatively different regimes for an ultrarelativistic electron  moving along the wave propagation direction: (a) $\xi_1\gg 1$ and $\xi_2\lesssim 1$, and (b) $\xi_1\gg\xi_2\gg1$, where $\xi_i$ ($i=1,2$) is the laser field parameter for the $i^{th}$ laser field, see Fig.~\ref{fig:illustration}. The analytical treatment employs an approximation for the classical trajectory based on the small parameter $\xi_2/\xi_1$.
The results are validated by a full numerical calculation.
A significant deviation from the LCFA is demonstrated in the high energy domain of the radiation spectra for both regimes even for strong laser fields, due to appearance of small characteristic time scales in the electron trajectory.
In case (a), the LCFA violation stems from oscillations of the trajectory within the $1/\gamma$-cone, resembling the one observed in a monochromatic plane wave.
In case (b), however, it results from the fact that the particle leaves the $1/\gamma$-cone more rapidly than predicted by the LCFA ($t_f<t_f^L$). Accordingly, the corresponding spectrum is broad and does not feature harmonics.  The same mechanism is also encountered in the interaction of an electron and an ultrashort pulse, where the new time scale is caused by the rapid changing of the pulse profile.


The quantum radiation process in a strong laser field is characterized by the classical nonlinear parameter $\xi$, and the quantum nonlinear parameter  $\chi= e\sqrt{-(F^{\mu \nu} P_{\nu})^2}/m^3$.
According to the Baier-Katkov approach, the radiation spectrum reads
\cite{Baier_b_1994}:
\begin{equation}
  \label{eq:bk_intensity}
    dI = \frac{\alpha}{(2 \pi)^2 \tau}   \left[
    -\frac{{\varepsilon'}^2 + {\varepsilon}^2 }{2  {\varepsilon}'^2}
    |\mathcal{T}_{\mu}|^2 +
    \frac{m ^2 \omega^2}{2 {\varepsilon'}^2 {\varepsilon}^2}
    |\mathcal{I}|^2
    \right] d^3\textbf{k} \,,
\end{equation}
where $\mathcal{I}
\equiv \int_{-\infty}^{\infty} e^{i \psi} \, \mathrm{d}t$ and
$\mathcal{T}_{\mu} \equiv \int_{-\infty}^{\infty} v_{\mu}(t) \, e^{i
  \psi} \, \mathrm{d}t$ with $\psi \equiv  \frac{\varepsilon }{\varepsilon'}k \cdot x(t)$ being the emission phase and $x_{\mu}(t)$, $k_{\mu}=(\omega,\textbf{k})$, $v_\mu(t)$ the four-vectors of the electron coordinate, the photon momentum and the velocity, respectively. $\tau$ is the pulse duration, $\varepsilon$ is the electron energy in the field and $\varepsilon' = \varepsilon - \omega$.

The CPW consists of two circularly polarized beams,  represented by the vector-potentials $A_1(x,t) = m \xi_1  [ \cos (k_1 \cdot x)
  e_x + \sin (k_1 \cdot x) e_y  ]$ and
$A_2(x,t) = m \xi_2  [ \cos (k_2 \cdot x) e_x + \sin (k_2
  \cdot x) e_y  ]$, with the four-wave-vectors $k_1 = (\omega_0,0,0,\omega_0)$
and $k_2 = (\omega_0,0,0,-\omega_0)$, respectively, with the optical frequency $\omega_0=1.55eV$, and $e_x = (0,1,0,0), e_y = (0,0,1,0)$. The classical Lorentz equation  is not solvable analytically for a general CPW
field.  We find an approximate solution for the electron trajectory imposing the conditions: $ \xi_1 \xi_2 \ll \gamma^2$, $ \xi_1 \gg \xi_2$, and with a vanishing asymptotic transverse momentum $p_{\bot}=0$:
\begin{eqnarray}
x(t) &=&[(\xi_1/\omega_1)\sin (\omega_1 t)+(\xi_2/\omega_2)\sin (\omega_2 t)]/\gamma, \nonumber\\
y(t) &=&[(\xi_1/\omega_1)\cos (\omega_1 t)+(\xi_2/\omega_2)\cos (\omega_2 t)]/\gamma,\nonumber\\
z(t)&=& \bar{v}_z t + 2 \omega_0  \xi_1 \xi_2 /(\gamma^2 \Delta \omega)\sin (\Delta \omega t ), \label{eq:xyz}
\end{eqnarray}
where $ \omega_1 \equiv \omega_0 \left( 1 - \bar{v}_z
\right), \omega_2 \equiv \omega_0 \left( 1 + \bar{v}_z
\right)$ and $\Delta \omega = \omega_1-\omega_2$. $\bar{v}_z$ is the average velocity of the electron which is copropagating with the $\xi_1$-beam.
The effective mass for the electron in the field is $m_* \approx m \sqrt{1+\xi_1^2+\xi_2^2}$.
The energy of the electron in the external field is constant
$\varepsilon \approx  \varepsilon_0 + m^2\xi_1^2/(\varepsilon_0-p_{z0})=m_*/\sqrt{1-\bar{v}_z} $, where $p^{\mu}_0$ is the asymptotic 4-momentum.

 \begin{figure}
  \begin{center}
  \includegraphics[width=0.5\textwidth]{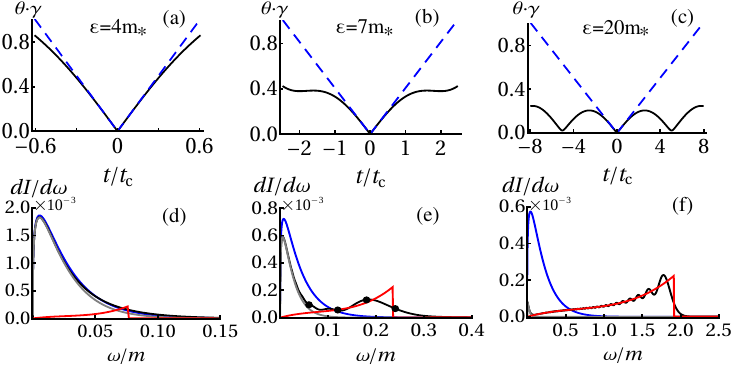}
  \caption{ Results for $\xi_1=20$ and $\xi_2=0.1$ with three
    different energies $\varepsilon=4m_*$ (first column),
    $\varepsilon=7m_*$ (second) and $\varepsilon=20m_*$ (third),
    respectively.   (a-c)  $\theta(t)$ in the vicinity of $t_0$
    when $\chi$ is maximal; dashed lines are $\theta_L(t)$ for LCFA, in this regime always $\theta(t)<\theta_L(t)$;
        (d-f) Radiation spectra;
           black lines are the exact spectra based on the analytical trajectory and blue lines are LCFA spectra with the time-dependent $\chi(t)$; gray and red
    lines are the spectra for a single beam with $\xi_1$ and $\xi_2$ respectively (the electron energy in the case of a single beam is chosen to be the same as for CPW in total). For comparison, the exact spectrum based on a fully numerical trajectory is also shown as the black dots in case (e).
        }
  \label{fig:lowxi}
  \end{center}
\end{figure}
The emission spectrum via Eqs.~\eqref{eq:bk_intensity}-\eqref{eq:xyz} is obtained in terms of Bessel functions, see  \cite{Supp}. The phase $\psi$ is an essential parameter determining the emitted radiation.
For the analysis of the radiation formation and its deviation from LCFA,
we approximate the trajectory in a short interval around time $t$,  defining the angle $\theta(t')$  of the velocity $\textbf{v}(t')$ with respect to $\textbf{v}(t) $:
 \begin{equation}
  \label{eq:bk_phase}
  \begin{split}
    \psi 
         \approx  \frac{\omega \varepsilon}{\varepsilon'} \left[ \frac{t}{2 \gamma^2} + \frac{v^2}{2}
                \int^t{dt'} \theta^2(t') \right] .
  \end{split}
\end{equation}
In LCFA \cite{Katkov_1968}, the angle is growing linearly  with time, $\theta_{L}(t)=m \chi t / \gamma^2$. Since the main contribution to the emission originates from $\theta \lesssim 1 / \gamma$, the formation time in LCFA is $t_f^L = 2 \gamma / (m \chi)$, which determines the typical energy of the emitted photon from the condition $\psi \sim 1$. Moreover, the formation time determines the LCFA applicability via $t_f^L\ll t_c$. In CPW  the characteristic time of the electron trajectory is \cite{Supp}:
\begin{equation}
  \label{eq:t_f}
    t_c= \chi(t)/\sqrt{\omega_1^2
      \chi_1^2 + \omega_2^2 \chi_2^2 - 2 \omega_1 \omega_2\chi_1 \chi_2
      \sin \left( \Delta \omega t \right) } \approx \frac{\chi(t)}{\chi_2 \omega_2}\nonumber
\end{equation}
where $\chi_1=\frac{\omega_1}{m} \xi_1 \gamma$ and $\chi_2=\frac{\omega_2}{m}\xi_2\gamma$ are the quantum parameters corresponding to the first and second beam alone, and
\begin{equation}
  \label{eq:chi_t}
  \chi(t) \approx \sqrt{\chi_1^2 + \chi_2^2 - 2 \chi_1 \chi_2 \cos \left( \Delta \omega t \right) }\,.
\end{equation}
Thus, the LCFA condition in the CPW setup is
\begin{equation}
  \label{eq:tr_tf_cpw}
  \frac{t_f^L}{t_c} \approx \frac{2 \gamma \omega_2 \chi_2}{m \chi^2(t) }\ll 1\,.
\end{equation}
In a single laser wave
it reduces to the familiar result $t_f^L/t_c \sim 1/ \xi$.
 In the following, the emitted spectrum is examined for two different parameter regimes with low $\xi_2$($ < 1$) and high $\xi_2$ ($ > 1$).

\textit{Low $\xi_2$ case}.  The radiation spectra corresponding to $\xi_1=20$ and $\xi_2=0.1$ are shown in Fig.~\ref{fig:lowxi}. We consider three electron energies $\varepsilon = 4m_* , 7m_*$ and $20m_*$, yielding different dynamics.  In the case of
$\varepsilon=4m_*$, the $\xi_1$-beam dominates the dynamics  because $\chi_1>\chi_2$.  As $\xi_1 \gg 1$, the angle $\theta(t)$ grows linearly within the $1/\gamma$-cone like in the LCFA case, see Fig.~\ref{fig:lowxi}(a).  Hence, the emitted spectrum
in Fig.~\ref{fig:lowxi}(d)
coincides with the LCFA result,  and
is close to the spectrum of a single beam with $\xi_1$.

For the high energy $\varepsilon=20m_*$,
 the $\xi_2$-beam is dominant and therefore  $t_f^L/t_c \approx 1/\xi_2 =10$, and the
angle  oscillates inside the $1/\gamma$-cone
(Fig.~\ref{fig:lowxi}(c)).
The radiation is similar to the case of the electron motion in a single $\xi_2$ with a renormalized energy due to the influence of the $\xi_1$-beam (Fig.~\ref{fig:lowxi}(f)). The typical energy of the emitted photon derived from $\psi\sim1$, is $ \omega=2 \gamma^2 \omega_2 \approx 4\gamma^2 \omega$.

Most interesting is the case of the intermediate energy $\varepsilon=7m_*$. Both beams influence
the dynamics as the quantum parameters are comparable, $\chi_1 \approx \chi_2$.
The ratio $t_f^L/t_c$ oscillates in time and is larger than unity,
which indicates the deviation from LCFA.  This can be seen from
Fig.~\ref{fig:lowxi}(b) where several $t_c$ periods contribute to the radiation during the electron oscillation within $1/\gamma$-cone,  see the  schematic trajectory in
Fig.~\ref{fig:illustration}(a). Note that in this regime $\theta(t)$ is always and mostly substantially smaller than the value via LCFA. The  spectrum (Fig.~\ref{fig:lowxi}(e)) reveals the qualitative deviations from the LCFA
result, as well as from those in single $\xi_1$ or $\xi_2$ beams.
This demonstrates the first evidence that even in a strong laser field
($\xi_1\gg 1$),  LCFA can completely underestimate the emission in the
high energy domain.

\begin{figure}
  \begin{center}
  \includegraphics[width=0.5\textwidth]{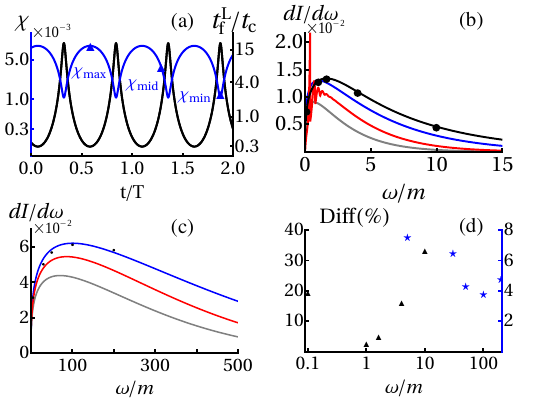}
  \caption{High $\xi_2$ case: (a) instantaneous $\chi(t)$ and
    $t_f^L/t_c$ vs time with
    $\mathrm{T}=2\pi/\omega_0$, $\xi_1=100$ and $\xi_2=2$.
Radiation spectrum: (b) for $\xi_1=100$ and $\xi_2=2$ ;  (c)  for  $\xi_1=500$ and $\xi_2=10$;  (d) the relative
    difference between the exact spectrum (black) and the LCFA one
    (blue), respectively, for the case of (b) (triangle) and for (c)
    (star).   The electron energy is $\varepsilon=4m_*$. The  color code is similar to
    Fig.~\ref{fig:lowxi}.}
  \label{fig:highxi1}
  \end{center}
\end{figure}

\textit{High $\xi_2$ case.}  We consider here $\xi_1=100$ and $\xi_2=2$ and demonstrate the LCFA can fail even though both lasers are strong.
Nontrivial radiation spectra are found when $\chi_1\sim \chi_2$, otherwise the spectrum approaches the corresponding single beam spectrum at $\chi_1 \gg \chi_2$ or vice
versa \cite{Supp}. We choose  $\varepsilon=4m_*$ so that $\chi_1\sim \chi_2 \sim 10^{-3}$.
From Fig.~\ref{fig:highxi1}(a) one can see that
$t_f^L/t_c$ is far above unity in the vicinity of the smallest $\chi(t)$
regime. This corresponds to the appearance of the smallest typical time scale in the trajectory ($t_3=\chi_{min}/(\chi_2 \omega_2) \ll 1/\omega_2=t_2$) as illustrated in Fig.~\ref{fig:illustration}(b), and leads to deviation of the spectrum from
the LCFA prediction, see Fig.~\ref{fig:highxi1}(b).
This deviation takes place in the entire energy domain, although the small time scale $t_3$  appears in parts
of the trajectory with low $\chi$ values.
Note that the emission in a single wave $\xi_1$ or $\xi_2$ here (red and gray lines) can be
described rather well with  LCFA  except for the
harmonic structure in the low energy region for $\xi_2=2$.

\begin{figure}[b]
  \begin{center}
  \includegraphics[width=0.5\textwidth]{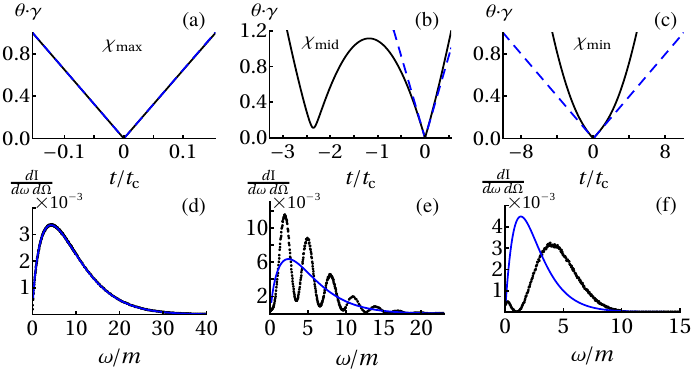}
  \caption{(a-c)  the angle $\theta(t)$ at different time
    points with different $\chi$ as shown in
    Fig.~\ref{fig:highxi1}(a).  The black and blue curves are for the
    exact trajectory and the LCFA estimation, respectively;
    $t_c=0.143 T$ for (a), $t_c=0.0763 T$ for (b), and
    $t_c=0.0179 T$ for (c);   (d-f) Angular
    resolved spectra for emitted direction
    corresponding to minimal, intermediate and maximal values of
    $\chi$.  A $\sin^2$-envelope for the laser pulse is applied in numerical calculations with 2-cycles up and 2-cycles down; the other parameters are the same as in Fig~\ref{fig:highxi1}.  }
  \label{fig:highxi2}
  \end{center}
\end{figure}

We analyze the deviation of radiation from LCFA by calculating the spectra
for three different directions (see Fig.~\ref{fig:highxi2}), corresponding to different $\chi(t) $ shown in
Fig.~\ref{fig:highxi1}(a). The exact angular resolved spectrum was
obtained by numerically solving for the trajectory and integrating Eq.~\eqref{eq:bk_intensity}. For clarity, we
avoid the  interference by choosing a short pulse
with only four cycles for $\xi_1$. From Fig.~\ref{fig:highxi2}(a),(d) one notes
that in the vicinity of $\chi_{max}$ the time-dependent angle
$\theta(t)$  and the spectrum agree well with the LCFA prediction, because
at this point  $t_f^L/t_\mathrm{c}=0.3$ [cf. the emission at $t_2$ in the scatch Fig~\ref{fig:illustration}(b)].
Around $\chi_{mid}$, however, the
particle oscillates in the $1/\gamma$-cone, as presented
in Fig.~\ref{fig:highxi2}(b). Accordingly, the emitted spectrum does
not agree with  LCFA but rather features harmonic structure,
see Fig.~\ref{fig:highxi2}(e).  One should notice that the fundamental
frequency of the harmonics $\sim 1/t_c$ is considerably larger than both
$\omega_1$ and $\omega_2$.  An extraordinary behavior emerges near
$\chi_{min}$. The angle $\theta(t)$ shown in Fig.~\ref{fig:highxi2}(c) is increasing more rapidly than $\theta_L(t)$. Namely, the deviation does not stem from oscillations within the $1/\gamma$ cone but from the fact that the particle exits it much quicker as compared to the LCFA estimation [cf. the emission at $t_3$ in the schematic of Fig~\ref{fig:illustration}(b)]. This accounts for the fact that the spectrum presented in
Fig.~\ref{fig:highxi2}(f) is broad and smooth, as opposed to the harmonic structure in
Fig.~\ref{fig:highxi2}(e).
This new class of LCFA violation is qualitatively distinct from the one observed in a monochromatic plane wave and is determined by the condition $t_f<t^L_f$, where the real formation time scale corresponds to $\theta(t_f) \sim 1/\gamma$.
The typical emitted photon energy may be estimated using $t_f$ and  the $\psi \sim 1$ condition
\begin{equation}
\omega \simeq  \frac{2 m \gamma^2 }{ m t_f/2 \Theta + 2 \gamma } \, ,
\label{eq:emit_energy}
\end{equation}
where $\Theta=1+ (\gamma^2/t_f) \int\displaylimits_{t_f} \theta^2(t') dt'$. As the second term in the latter is $\sim 1$, we obtain an  estimation
 $\omega\simeq 2 m \gamma^2 /( m t_f + 2 \gamma )$.
In the LCFA case, $t_f^L=2\gamma/(m\chi)$, so that
the familiar result $ \omega \sim
\varepsilon \chi/(1+\chi) $ is recovered.  One  notes from
Fig.~\ref{fig:highxi2}(c) that $t_f\ll t_f^L$.
Consequently, the typical energy of the emitted photons in Fig.~\ref{fig:highxi2}(f)
is considerably larger than the 
LCFA one
 and agrees with the estimation $\omega  \approx 4.2 m$ from
Eq.~\eqref{eq:emit_energy}.  Furthermore, the spectra shown in
Fig.~\ref{fig:highxi2}(d, e, f) have  similar amplitudes,
explaining the high energy deviations  in
Fig.~\ref{fig:highxi1}(b).

The described  deviation from LCFA persists also at higher laser intensities, as shown in the case of $\xi_1=500$ and $\xi_2=10$ in Fig.~\ref{fig:highxi1}(c), albeit
the emission of each single beam can be well represented by LCFA. The evaluation of the analytical expression for this case is unpractical, involving summing extremely high numbers of
Bessel functions. Hence, only the numerical calculation is presented.
The deviation here is weaker as compared to  Fig.~\ref{fig:highxi1}(b) (see panel (d)), since the relative part of the trajectory with $t_f^L>t_c$ is shorter, see \cite{Supp}.
This discrepancy, though quantitative rather than qualitative, bears significance since multiple photon emissions are probable in ultrastrong fields and even a minor difference in each single emission will be accumulated.



\begin{figure}
  \begin{center}
  \includegraphics[width=0.5\textwidth]{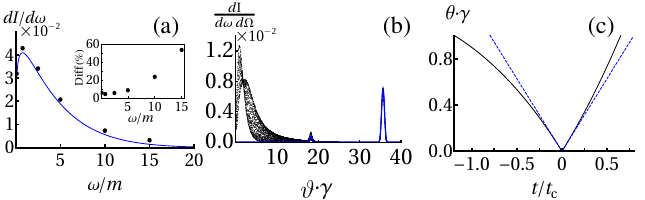}
  \caption{Radiation in an ultrashort laser pulse:  (a) The spectrum  with
     LCFA  (blue) and the exact numerical calculation (black). The inset is the relative difference between the exact and LCFA spectra;   (b) Angular resolved
    spectrum with a fixed azimuthal direction $\varphi=3\pi/4$ and for  $\omega=10m$; (c)  $\theta(t)$
    for $t_0=-0.65\mathrm{T}$ corresponding to $\gamma \vartheta=1.25$ in panel(b). The laser beam has a Gaussian profile with a standard deviation $\sigma=0.5$ and the peak intensity is $\xi=50$ at $t=0$. The waist size is
    $w_0=3\lambda_0$, and the frequency is $\omega_0=1.55eV$. The electron with energy of 100$m$ counterpropagates with respect to the laser beam.}
      \label{fig:short_focuse}
  \end{center}
\end{figure}

An analogous deviation from LCFA due to the emergence of a small characteristic time scale in the electron trajectory can also happen in a simpler field configuration.  
We have analyzed the radiation emitted by an ultrarelativistic electron colliding with a single ultrashort laser pulse \cite{Supp}, where the characteristic time scale of the electron trajectory is shaped not only by the central frequency of the laser wave, but also by the time-envelope of the laser pulse.
The comparison of the exact radiation spectra calculated numerically with that of LCFA is presented in Fig.~\ref{fig:short_focuse}.

Surprisingly, even though $\xi\gg 1$, a difference between the LCFA prediction and the exact result exists through the entire spectrum, including high energies (see Fig.~\ref{fig:short_focuse}(a)).
To seek the reason of the deviation, the angular
resolved spectrum with a fixed azimuthal direction $\varphi=3\pi/4$ for
the emitted energy $\omega=10m$ is displayed in Fig.~\ref{fig:short_focuse}(b).  One can see that the main difference 
corresponds to low $\vartheta$ value, namely to the beginning and
the end of the trajectory. The corresponding characteristic time 
$t_c=0.0613T$ is rather small as compared to the laser period and thus $t_f^L/t_c=1.59>1$ \cite{Supp}. Since $t_c \equiv |\vec{F}|/|\dot{\vec{F}}|$ this can be understood as $\dot{\vec{F}}$ is large in the beginning and the end of the ultrashort pulse.
Moreover, the direction of motion is changing rapidly, see  Fig.~\ref{fig:short_focuse}(c), in accordance with the new class of violation.  
Here, similar to the CPW case of Fig.~\ref{fig:highxi2}(f),
the rise and the fall of the ultrashort laser pulse
can influence the emission in all spectral range, even though the
quantum parameter $\chi(t)$ is relatively small in this region.

\textit{Conclusion.} We have studied the validity of LCFA for the description of an ultrarelativistic electron radiation in ultrastrong laser fields, in a CPW multiple beam configuration and in an ultrashort laser pulse.  In both cases deviations from LCFA are observed in the radiation spectra, in particular in the high photon energy domain. 
This deviation results from the emergence of extra small time scales of the electron trajectory, which are comparable to the radiation formation time $t_f\gtrsim t_c$. 
The shortness of the emerged time scales is the reason for the disturbance of the high-energy spectral region.
Moreover, we identify a novel class of LCFA violation, where the radiating electron leaves the $1/\gamma$-cone much faster than along the approximate LCFA trajectory (so that $t_f < t_f^L$).
Our discussion has a direct implication for the calculation of electron radiation in laser-plasma interaction. In the latter multiple beam configuration may arise, when a strong laser wave impinging on the target is reflected from the critical density, or when a counterpropagating  plasma wave is induced.  Our results indicate that LCFA should be applied with particular care for complex field configurations in Monte Carlo simulations of the laser-matter interaction.

Q.Z.L and E.R. contributed equally to the work, to numerical and analytical calculations, respectively.
E.R. acknowledges partial support from the Alexander von Humboldt Foundation.

\bibliography{criteria_lcfa}

\begin{thebibliography}{43}%
\makeatletter
\providecommand \@ifxundefined [1]{%
 \@ifx{#1\undefined}
}%
\providecommand \@ifnum [1]{%
 \ifnum #1\expandafter \@firstoftwo
 \else \expandafter \@secondoftwo
 \fi
}%
\providecommand \@ifx [1]{%
 \ifx #1\expandafter \@firstoftwo
 \else \expandafter \@secondoftwo
 \fi
}%
\providecommand \natexlab [1]{#1}%
\providecommand \enquote  [1]{``#1''}%
\providecommand \bibnamefont  [1]{#1}%
\providecommand \bibfnamefont [1]{#1}%
\providecommand \citenamefont [1]{#1}%
\providecommand \href@noop [0]{\@secondoftwo}%
\providecommand \href [0]{\begingroup \@sanitize@url \@href}%
\providecommand \@href[1]{\@@startlink{#1}\@@href}%
\providecommand \@@href[1]{\endgroup#1\@@endlink}%
\providecommand \@sanitize@url [0]{\catcode `\\12\catcode `\$12\catcode
  `\&12\catcode `\#12\catcode `\^12\catcode `\_12\catcode `\%12\relax}%
\providecommand \@@startlink[1]{}%
\providecommand \@@endlink[0]{}%
\providecommand \url  [0]{\begingroup\@sanitize@url \@url }%
\providecommand \@url [1]{\endgroup\@href {#1}{\urlprefix }}%
\providecommand \urlprefix  [0]{URL }%
\providecommand \Eprint [0]{\href }%
\providecommand \doibase [0]{http://dx.doi.org/}%
\providecommand \selectlanguage [0]{\@gobble}%
\providecommand \bibinfo  [0]{\@secondoftwo}%
\providecommand \bibfield  [0]{\@secondoftwo}%
\providecommand \translation [1]{[#1]}%
\providecommand \BibitemOpen [0]{}%
\providecommand \bibitemStop [0]{}%
\providecommand \bibitemNoStop [0]{.\EOS\space}%
\providecommand \EOS [0]{\spacefactor3000\relax}%
\providecommand \BibitemShut  [1]{\csname bibitem#1\endcsname}%
\let\auto@bib@innerbib\@empty
\bibitem [{\citenamefont {Mourou}\ \emph {et~al.}(2006)\citenamefont {Mourou},
  \citenamefont {Tajima},\ and\ \citenamefont {Bulanov}}]{Mourou_2006}%
  \BibitemOpen
  \bibfield  {author} {\bibinfo {author} {\bibfnamefont {G~A}\ \bibnamefont
  {Mourou}}, \bibinfo {author} {\bibfnamefont {T}~\bibnamefont {Tajima}}, \
  and\ \bibinfo {author} {\bibfnamefont {S~V}\ \bibnamefont {Bulanov}},\
  }\bibfield  {title} {\enquote {\bibinfo {title} {{Optics in the relativistic
  regime}},}\ }\href@noop {} {\bibfield  {journal} {\bibinfo  {journal} {Rev.
  Mod. Phys.}\ }\textbf {\bibinfo {volume} {78}},\ \bibinfo {pages} {309--371}
  (\bibinfo {year} {2006})}\BibitemShut {NoStop}%
\bibitem [{\citenamefont {Marklund}\ and\ \citenamefont
  {Shukla}(2006)}]{Marklund_2006}%
  \BibitemOpen
  \bibfield  {author} {\bibinfo {author} {\bibfnamefont {M.}~\bibnamefont
  {Marklund}}\ and\ \bibinfo {author} {\bibfnamefont {P.~K.}\ \bibnamefont
  {Shukla}},\ }\bibfield  {title} {\enquote {\bibinfo {title} {Nonlinear
  collective effects in photon-photon and photon-plasma interactions},}\
  }\href@noop {} {\bibfield  {journal} {\bibinfo  {journal} {Rev. Mod. Phys.}\
  }\textbf {\bibinfo {volume} {78}},\ \bibinfo {pages} {591} (\bibinfo {year}
  {2006})}\BibitemShut {NoStop}%
\bibitem [{\citenamefont {{Di Piazza}}\ \emph {et~al.}(2012)\citenamefont {{Di
  Piazza}}, \citenamefont {M\"uller}, \citenamefont {Hatsagortsyan},\ and\
  \citenamefont {Keitel}}]{RMP_2012}%
  \BibitemOpen
  \bibfield  {author} {\bibinfo {author} {\bibfnamefont {A.}~\bibnamefont {{Di
  Piazza}}}, \bibinfo {author} {\bibfnamefont {C.}~\bibnamefont {M\"uller}},
  \bibinfo {author} {\bibfnamefont {K.~Z.}\ \bibnamefont {Hatsagortsyan}}, \
  and\ \bibinfo {author} {\bibfnamefont {C.~H.}\ \bibnamefont {Keitel}},\
  }\bibfield  {title} {\enquote {\bibinfo {title} {Extremely high-intensity
  laser interactions with fundamental quantum systems},}\ }\href@noop {}
  {\bibfield  {journal} {\bibinfo  {journal} {Rev. Mod. Phys.}\ }\textbf
  {\bibinfo {volume} {84}},\ \bibinfo {pages} {1177} (\bibinfo {year}
  {2012})}\BibitemShut {NoStop}%
\bibitem [{\citenamefont {Heinzl}(2012)}]{Heinzl_2012}%
  \BibitemOpen
  \bibfield  {author} {\bibinfo {author} {\bibfnamefont {T.}~\bibnamefont
  {Heinzl}},\ }\bibfield  {title} {\enquote {\bibinfo {title} {Strong-field qed
  and high-power lasers},}\ }\href@noop {} {\bibfield  {journal} {\bibinfo
  {journal} {Int. J. Mod. Phys. A}\ }\textbf {\bibinfo {volume} {27}},\
  \bibinfo {pages} {1260010} (\bibinfo {year} {2012})}\BibitemShut {NoStop}%
\bibitem [{\citenamefont {Dunne}(2014)}]{Dunne2014}%
  \BibitemOpen
  \bibfield  {author} {\bibinfo {author} {\bibfnamefont {G.V.}\ \bibnamefont
  {Dunne}},\ }\href@noop {} {\bibfield  {journal} {\bibinfo  {journal} {Eur.
  Phys. J. Spec. Top.}\ }\textbf {\bibinfo {volume} {223}},\ \bibinfo {pages}
  {1055} (\bibinfo {year} {2014})}\BibitemShut {NoStop}%
\bibitem [{\citenamefont {Turcu}\ \emph {et~al.}(2019)\citenamefont {Turcu},
  \citenamefont {Shen}, \citenamefont {Neely}, \citenamefont {Sarri},
  \citenamefont {Tanaka}, \citenamefont {McKenna}, \citenamefont {Mangles},
  \citenamefont {Yu}, \citenamefont {Luo}, \citenamefont {Zhu},\ and\
  \citenamefont {et~al.}}]{Turcu_2019}%
  \BibitemOpen
  \bibfield  {author} {\bibinfo {author} {\bibfnamefont {I.~C.~E.}\
  \bibnamefont {Turcu}}, \bibinfo {author} {\bibfnamefont {B.}~\bibnamefont
  {Shen}}, \bibinfo {author} {\bibfnamefont {D.}~\bibnamefont {Neely}},
  \bibinfo {author} {\bibfnamefont {G.}~\bibnamefont {Sarri}}, \bibinfo
  {author} {\bibfnamefont {K.~A.}\ \bibnamefont {Tanaka}}, \bibinfo {author}
  {\bibfnamefont {P.}~\bibnamefont {McKenna}}, \bibinfo {author} {\bibfnamefont
  {S.~P.~D.}\ \bibnamefont {Mangles}}, \bibinfo {author} {\bibfnamefont
  {T.-P.}\ \bibnamefont {Yu}}, \bibinfo {author} {\bibfnamefont
  {W.}~\bibnamefont {Luo}}, \bibinfo {author} {\bibfnamefont {X.-L.}\
  \bibnamefont {Zhu}}, \ and\ \bibinfo {author} {\bibnamefont {et~al.}},\
  }\bibfield  {title} {\enquote {\bibinfo {title} {Quantum electrodynamics
  experiments with colliding petawatt laser pulses},}\ }\href@noop {}
  {\bibfield  {journal} {\bibinfo  {journal} {High Power Laser Sci. Eng.}\
  }\textbf {\bibinfo {volume} {7}},\ \bibinfo {pages} {e10} (\bibinfo {year}
  {2019})}\BibitemShut {NoStop}%
\bibitem [{\citenamefont {Bula}\ \emph {et~al.}(1996)\citenamefont {Bula},
  \citenamefont {McDonald}, \citenamefont {Prebys}, \citenamefont {Bamber},
  \citenamefont {Boege}, \citenamefont {Kotseroglou}, \citenamefont
  {Melissinos}, \citenamefont {Meyerhofer}, \citenamefont {Ragg}, \citenamefont
  {Burke}, \citenamefont {Field}, \citenamefont {Horton-Smith}, \citenamefont
  {Odian}, \citenamefont {Spencer}, \citenamefont {Walz}, \citenamefont
  {Berridge}, \citenamefont {Bugg}, \citenamefont {Shmakov},\ and\
  \citenamefont {Weidemann}}]{Bula_1996}%
  \BibitemOpen
  \bibfield  {author} {\bibinfo {author} {\bibfnamefont {C.}~\bibnamefont
  {Bula}}, \bibinfo {author} {\bibfnamefont {K.~T.}\ \bibnamefont {McDonald}},
  \bibinfo {author} {\bibfnamefont {E.~J.}\ \bibnamefont {Prebys}}, \bibinfo
  {author} {\bibfnamefont {C.}~\bibnamefont {Bamber}}, \bibinfo {author}
  {\bibfnamefont {S.}~\bibnamefont {Boege}}, \bibinfo {author} {\bibfnamefont
  {T.}~\bibnamefont {Kotseroglou}}, \bibinfo {author} {\bibfnamefont {A.~C.}\
  \bibnamefont {Melissinos}}, \bibinfo {author} {\bibfnamefont {D.~D.}\
  \bibnamefont {Meyerhofer}}, \bibinfo {author} {\bibfnamefont
  {W.}~\bibnamefont {Ragg}}, \bibinfo {author} {\bibfnamefont {D.~L.}\
  \bibnamefont {Burke}}, \bibinfo {author} {\bibfnamefont {R.~C.}\ \bibnamefont
  {Field}}, \bibinfo {author} {\bibfnamefont {G.}~\bibnamefont {Horton-Smith}},
  \bibinfo {author} {\bibfnamefont {A.~C.}\ \bibnamefont {Odian}}, \bibinfo
  {author} {\bibfnamefont {J.~E.}\ \bibnamefont {Spencer}}, \bibinfo {author}
  {\bibfnamefont {D.}~\bibnamefont {Walz}}, \bibinfo {author} {\bibfnamefont
  {S.~C.}\ \bibnamefont {Berridge}}, \bibinfo {author} {\bibfnamefont {W.~M.}\
  \bibnamefont {Bugg}}, \bibinfo {author} {\bibfnamefont {K.}~\bibnamefont
  {Shmakov}}, \ and\ \bibinfo {author} {\bibfnamefont {A.~W.}\ \bibnamefont
  {Weidemann}},\ }\bibfield  {title} {\enquote {\bibinfo {title} {Observation
  of nonlinear effects in compton scattering},}\ }\href@noop {} {\bibfield
  {journal} {\bibinfo  {journal} {Phys. Rev. Lett.}\ }\textbf {\bibinfo
  {volume} {76}},\ \bibinfo {pages} {3116--3119} (\bibinfo {year}
  {1996})}\BibitemShut {NoStop}%
\bibitem [{\citenamefont {Burke}\ \emph {et~al.}(1997)\citenamefont {Burke},
  \citenamefont {Field}, \citenamefont {Horton-Smith}, \citenamefont {Spencer},
  \citenamefont {Walz}, \citenamefont {Berridge}, \citenamefont {Bugg},
  \citenamefont {Shmakov}, \citenamefont {Weidemann}, \citenamefont {Bula},
  \citenamefont {McDonald}, \citenamefont {Prebys}, \citenamefont {Bamber},
  \citenamefont {Boege}, \citenamefont {Koffas}, \citenamefont {Kotseroglou},
  \citenamefont {Melissinos}, \citenamefont {Meyerhofer}, \citenamefont
  {Reis},\ and\ \citenamefont {Ragg}}]{Burke_1997}%
  \BibitemOpen
  \bibfield  {author} {\bibinfo {author} {\bibfnamefont {D.~L.}\ \bibnamefont
  {Burke}}, \bibinfo {author} {\bibfnamefont {R.~C.}\ \bibnamefont {Field}},
  \bibinfo {author} {\bibfnamefont {G.}~\bibnamefont {Horton-Smith}}, \bibinfo
  {author} {\bibfnamefont {J.~E.}\ \bibnamefont {Spencer}}, \bibinfo {author}
  {\bibfnamefont {D.}~\bibnamefont {Walz}}, \bibinfo {author} {\bibfnamefont
  {S.~C.}\ \bibnamefont {Berridge}}, \bibinfo {author} {\bibfnamefont {W.~M.}\
  \bibnamefont {Bugg}}, \bibinfo {author} {\bibfnamefont {K.}~\bibnamefont
  {Shmakov}}, \bibinfo {author} {\bibfnamefont {A.~W.}\ \bibnamefont
  {Weidemann}}, \bibinfo {author} {\bibfnamefont {C.}~\bibnamefont {Bula}},
  \bibinfo {author} {\bibfnamefont {K.~T.}\ \bibnamefont {McDonald}}, \bibinfo
  {author} {\bibfnamefont {E.~J.}\ \bibnamefont {Prebys}}, \bibinfo {author}
  {\bibfnamefont {C.}~\bibnamefont {Bamber}}, \bibinfo {author} {\bibfnamefont
  {S.~J.}\ \bibnamefont {Boege}}, \bibinfo {author} {\bibfnamefont
  {T.}~\bibnamefont {Koffas}}, \bibinfo {author} {\bibfnamefont
  {T.}~\bibnamefont {Kotseroglou}}, \bibinfo {author} {\bibfnamefont {A.~C.}\
  \bibnamefont {Melissinos}}, \bibinfo {author} {\bibfnamefont {D.~D.}\
  \bibnamefont {Meyerhofer}}, \bibinfo {author} {\bibfnamefont {D.~A.}\
  \bibnamefont {Reis}}, \ and\ \bibinfo {author} {\bibfnamefont
  {W.}~\bibnamefont {Ragg}},\ }\bibfield  {title} {\enquote {\bibinfo {title}
  {Positron production in multiphoton light-by-light scattering},}\ }\href
  {\doibase 10.1103/PhysRevLett.79.1626} {\bibfield  {journal} {\bibinfo
  {journal} {Phys. Rev. Lett.}\ }\textbf {\bibinfo {volume} {79}},\ \bibinfo
  {pages} {1626} (\bibinfo {year} {1997})}\BibitemShut {NoStop}%
\bibitem [{\citenamefont {Cole}\ \emph {et~al.}(2018)\citenamefont {Cole},
  \citenamefont {Behm}, \citenamefont {Gerstmayr}, \citenamefont {Blackburn},
  \citenamefont {Wood}, \citenamefont {Baird}, \citenamefont {Duff},
  \citenamefont {Harvey}, \citenamefont {Ilderton}, \citenamefont {Joglekar},
  \citenamefont {Krushelnick}, \citenamefont {Kuschel}, \citenamefont
  {Marklund}, \citenamefont {McKenna}, \citenamefont {Murphy}, \citenamefont
  {Poder}, \citenamefont {Ridgers}, \citenamefont {Samarin}, \citenamefont
  {Sarri}, \citenamefont {Symes}, \citenamefont {Thomas}, \citenamefont
  {Warwick}, \citenamefont {Zepf}, \citenamefont {Najmudin},\ and\
  \citenamefont {Mangles}}]{Cole_2018}%
  \BibitemOpen
  \bibfield  {author} {\bibinfo {author} {\bibfnamefont {J.~M.}\ \bibnamefont
  {Cole}}, \bibinfo {author} {\bibfnamefont {K.~T.}\ \bibnamefont {Behm}},
  \bibinfo {author} {\bibfnamefont {E.}~\bibnamefont {Gerstmayr}}, \bibinfo
  {author} {\bibfnamefont {T.~G.}\ \bibnamefont {Blackburn}}, \bibinfo {author}
  {\bibfnamefont {J.~C.}\ \bibnamefont {Wood}}, \bibinfo {author}
  {\bibfnamefont {C.~D.}\ \bibnamefont {Baird}}, \bibinfo {author}
  {\bibfnamefont {M.~J.}\ \bibnamefont {Duff}}, \bibinfo {author}
  {\bibfnamefont {C.}~\bibnamefont {Harvey}}, \bibinfo {author} {\bibfnamefont
  {A.}~\bibnamefont {Ilderton}}, \bibinfo {author} {\bibfnamefont {A.~S.}\
  \bibnamefont {Joglekar}}, \bibinfo {author} {\bibfnamefont {K.}~\bibnamefont
  {Krushelnick}}, \bibinfo {author} {\bibfnamefont {S.}~\bibnamefont
  {Kuschel}}, \bibinfo {author} {\bibfnamefont {M.}~\bibnamefont {Marklund}},
  \bibinfo {author} {\bibfnamefont {P.}~\bibnamefont {McKenna}}, \bibinfo
  {author} {\bibfnamefont {C.~D.}\ \bibnamefont {Murphy}}, \bibinfo {author}
  {\bibfnamefont {K.}~\bibnamefont {Poder}}, \bibinfo {author} {\bibfnamefont
  {C.~P.}\ \bibnamefont {Ridgers}}, \bibinfo {author} {\bibfnamefont {G.~M.}\
  \bibnamefont {Samarin}}, \bibinfo {author} {\bibfnamefont {G.}~\bibnamefont
  {Sarri}}, \bibinfo {author} {\bibfnamefont {D.~R.}\ \bibnamefont {Symes}},
  \bibinfo {author} {\bibfnamefont {A.~G.~R.}\ \bibnamefont {Thomas}}, \bibinfo
  {author} {\bibfnamefont {J.}~\bibnamefont {Warwick}}, \bibinfo {author}
  {\bibfnamefont {M.}~\bibnamefont {Zepf}}, \bibinfo {author} {\bibfnamefont
  {Z.}~\bibnamefont {Najmudin}}, \ and\ \bibinfo {author} {\bibfnamefont
  {S.~P.~D.}\ \bibnamefont {Mangles}},\ }\bibfield  {title} {\enquote {\bibinfo
  {title} {Experimental evidence of radiation reaction in the collision of a
  high-intensity laser pulse with a laser-wakefield accelerated electron
  beam},}\ }\href@noop {} {\bibfield  {journal} {\bibinfo  {journal} {Phys.
  Rev. X}\ }\textbf {\bibinfo {volume} {8}},\ \bibinfo {pages} {011020}
  (\bibinfo {year} {2018})}\BibitemShut {NoStop}%
\bibitem [{\citenamefont {Poder}\ \emph {et~al.}(2018)\citenamefont {Poder},
  \citenamefont {Tamburini}, \citenamefont {Sarri}, \citenamefont {Di~Piazza},
  \citenamefont {Kuschel}, \citenamefont {Baird}, \citenamefont {Behm},
  \citenamefont {Bohlen}, \citenamefont {Cole}, \citenamefont {Corvan},
  \citenamefont {Duff}, \citenamefont {Gerstmayr}, \citenamefont {Keitel},
  \citenamefont {Krushelnick}, \citenamefont {Mangles}, \citenamefont
  {McKenna}, \citenamefont {Murphy}, \citenamefont {Najmudin}, \citenamefont
  {Ridgers}, \citenamefont {Samarin}, \citenamefont {Symes}, \citenamefont
  {Thomas}, \citenamefont {Warwick},\ and\ \citenamefont {Zepf}}]{Poder_2018}%
  \BibitemOpen
  \bibfield  {author} {\bibinfo {author} {\bibfnamefont {K.}~\bibnamefont
  {Poder}}, \bibinfo {author} {\bibfnamefont {M.}~\bibnamefont {Tamburini}},
  \bibinfo {author} {\bibfnamefont {G.}~\bibnamefont {Sarri}}, \bibinfo
  {author} {\bibfnamefont {A.}~\bibnamefont {Di~Piazza}}, \bibinfo {author}
  {\bibfnamefont {S.}~\bibnamefont {Kuschel}}, \bibinfo {author} {\bibfnamefont
  {C.~D.}\ \bibnamefont {Baird}}, \bibinfo {author} {\bibfnamefont
  {K.}~\bibnamefont {Behm}}, \bibinfo {author} {\bibfnamefont {S.}~\bibnamefont
  {Bohlen}}, \bibinfo {author} {\bibfnamefont {J.~M.}\ \bibnamefont {Cole}},
  \bibinfo {author} {\bibfnamefont {D.~J.}\ \bibnamefont {Corvan}}, \bibinfo
  {author} {\bibfnamefont {M.}~\bibnamefont {Duff}}, \bibinfo {author}
  {\bibfnamefont {E.}~\bibnamefont {Gerstmayr}}, \bibinfo {author}
  {\bibfnamefont {C.~H.}\ \bibnamefont {Keitel}}, \bibinfo {author}
  {\bibfnamefont {K.}~\bibnamefont {Krushelnick}}, \bibinfo {author}
  {\bibfnamefont {S.~P.~D.}\ \bibnamefont {Mangles}}, \bibinfo {author}
  {\bibfnamefont {P.}~\bibnamefont {McKenna}}, \bibinfo {author} {\bibfnamefont
  {C.~D.}\ \bibnamefont {Murphy}}, \bibinfo {author} {\bibfnamefont
  {Z.}~\bibnamefont {Najmudin}}, \bibinfo {author} {\bibfnamefont {C.~P.}\
  \bibnamefont {Ridgers}}, \bibinfo {author} {\bibfnamefont {G.~M.}\
  \bibnamefont {Samarin}}, \bibinfo {author} {\bibfnamefont {D.~R.}\
  \bibnamefont {Symes}}, \bibinfo {author} {\bibfnamefont {A.~G.~R.}\
  \bibnamefont {Thomas}}, \bibinfo {author} {\bibfnamefont {J.}~\bibnamefont
  {Warwick}}, \ and\ \bibinfo {author} {\bibfnamefont {M.}~\bibnamefont
  {Zepf}},\ }\bibfield  {title} {\enquote {\bibinfo {title} {Experimental
  signatures of the quantum nature of radiation reaction in the field of an
  ultraintense laser},}\ }\href@noop {} {\bibfield  {journal} {\bibinfo
  {journal} {Phys. Rev. X}\ }\textbf {\bibinfo {volume} {8}},\ \bibinfo {pages}
  {031004} (\bibinfo {year} {2018})}\BibitemShut {NoStop}%
\bibitem [{\citenamefont {Strickland}(2019)}]{Strickland_2019}%
  \BibitemOpen
  \bibfield  {author} {\bibinfo {author} {\bibfnamefont {D.}~\bibnamefont
  {Strickland}},\ }\bibfield  {title} {\enquote {\bibinfo {title} {{Nobel
  Lecture: Generating high-intensity ultrashort optical pulses}},}\ }\href@noop
  {} {\bibfield  {journal} {\bibinfo  {journal} {Rev. Mod. Phys.}\ }\textbf
  {\bibinfo {volume} {91}},\ \bibinfo {pages} {030502} (\bibinfo {year}
  {2019})}\BibitemShut {NoStop}%
\bibitem [{\citenamefont {Mourou}(2019)}]{Mourou_2019}%
  \BibitemOpen
  \bibfield  {author} {\bibinfo {author} {\bibfnamefont {G.}~\bibnamefont
  {Mourou}},\ }\bibfield  {title} {\enquote {\bibinfo {title} {{Nobel Lecture:
  Extreme light physics and application}},}\ }\href@noop {} {\bibfield
  {journal} {\bibinfo  {journal} {Rev. Mod. Phys.}\ }\textbf {\bibinfo {volume}
  {91}},\ \bibinfo {pages} {030501} (\bibinfo {year} {2019})}\BibitemShut
  {NoStop}%
\bibitem [{\citenamefont {Yoon}\ \emph {et~al.}(2019)\citenamefont {Yoon},
  \citenamefont {Jeon}, \citenamefont {Shin}, \citenamefont {Lee},
  \citenamefont {Lee}, \citenamefont {Choi}, \citenamefont {Kim}, \citenamefont
  {Sung}, ,\ and\ \citenamefont {Nam}}]{Yoon_2019}%
  \BibitemOpen
  \bibfield  {author} {\bibinfo {author} {\bibfnamefont {J.W.}\ \bibnamefont
  {Yoon}}, \bibinfo {author} {\bibfnamefont {C.}~\bibnamefont {Jeon}}, \bibinfo
  {author} {\bibfnamefont {J.}~\bibnamefont {Shin}}, \bibinfo {author}
  {\bibfnamefont {S.K.}\ \bibnamefont {Lee}}, \bibinfo {author} {\bibfnamefont
  {H.W.}\ \bibnamefont {Lee}}, \bibinfo {author} {\bibfnamefont {I.W.}\
  \bibnamefont {Choi}}, \bibinfo {author} {\bibfnamefont {H.T.}\ \bibnamefont
  {Kim}}, \bibinfo {author} {\bibfnamefont {J.H.}\ \bibnamefont {Sung}}, , \
  and\ \bibinfo {author} {\bibfnamefont {C.H.}\ \bibnamefont {Nam}},\
  }\href@noop {} {\bibfield  {journal} {\bibinfo  {journal} {Opt. Express}\
  }\textbf {\bibinfo {volume} {27}},\ \bibinfo {pages} {20412} (\bibinfo {year}
  {2019})}\BibitemShut {NoStop}%
\bibitem [{\citenamefont {{The Vulcan facility}}()}]{Vulcan1}%
  \BibitemOpen
  \bibfield  {author} {\bibinfo {author} {\bibnamefont {{The Vulcan
  facility}}},\ }\href@noop {} {}\bibinfo {howpublished}
  {\url{https://www.clf.stfc.ac.uk/Pages/ Vulcan-laser.aspx}}\BibitemShut
  {NoStop}%
\bibitem [{\citenamefont {{The Extreme Light Infrastructure (ELI)}}()}]{ELI}%
  \BibitemOpen
  \bibfield  {author} {\bibinfo {author} {\bibnamefont {{The Extreme Light
  Infrastructure (ELI)}}},\ }\href@noop {} {}\bibinfo {howpublished}
  {\url{http://www.eli-laser.eu/}}\BibitemShut {NoStop}%
\bibitem [{\citenamefont {{Exawatt Center for Extreme Light Stidies
  (XCELS)}}()}]{XCELS}%
  \BibitemOpen
  \bibfield  {author} {\bibinfo {author} {\bibnamefont {{Exawatt Center for
  Extreme Light Stidies (XCELS)}}},\ }\href@noop {} {}\bibinfo {howpublished}
  {\url{http://www.xcels.iapras.ru/}}\BibitemShut {NoStop}%
\bibitem [{\citenamefont {Wolkow}(1935)}]{Volkov_1935}%
  \BibitemOpen
  \bibfield  {author} {\bibinfo {author} {\bibfnamefont {D.~M.}\ \bibnamefont
  {Wolkow}},\ }\bibfield  {title} {\enquote {\bibinfo {title} {{Über eine
  Klasse von Lösungen der Diracschen Gleichung}},}\ }\href@noop {} {\bibfield
  {journal} {\bibinfo  {journal} {Z. Phys.}\ }\textbf {\bibinfo {volume}
  {94}},\ \bibinfo {pages} {250} (\bibinfo {year} {1935})}\BibitemShut
  {NoStop}%
\bibitem [{\citenamefont {{Di~Piazza}}(2014)}]{DiPiazza_2014}%
  \BibitemOpen
  \bibfield  {author} {\bibinfo {author} {\bibfnamefont {A.}~\bibnamefont
  {{Di~Piazza}}},\ }\bibfield  {title} {\enquote {\bibinfo {title}
  {Ultrarelativistic electron states in a general background electromagnetic
  field},}\ }\href@noop {} {\bibfield  {journal} {\bibinfo  {journal} {Phys.
  Rev. Lett.}\ }\textbf {\bibinfo {volume} {113}},\ \bibinfo {pages} {040402}
  (\bibinfo {year} {2014})}\BibitemShut {NoStop}%
\bibitem [{\citenamefont {Elkina}\ \emph {et~al.}(2011)\citenamefont {Elkina},
  \citenamefont {Fedotov}, \citenamefont {Kostyukov}, \citenamefont {Legkov},
  \citenamefont {Narozhny}, \citenamefont {Nerush},\ and\ \citenamefont
  {Ruhl}}]{Elkina_2011}%
  \BibitemOpen
  \bibfield  {author} {\bibinfo {author} {\bibfnamefont {N.~V.}\ \bibnamefont
  {Elkina}}, \bibinfo {author} {\bibfnamefont {A.~M.}\ \bibnamefont {Fedotov}},
  \bibinfo {author} {\bibfnamefont {I.~Yu.}\ \bibnamefont {Kostyukov}},
  \bibinfo {author} {\bibfnamefont {M.~V.}\ \bibnamefont {Legkov}}, \bibinfo
  {author} {\bibfnamefont {N.~B.}\ \bibnamefont {Narozhny}}, \bibinfo {author}
  {\bibfnamefont {E.~N.}\ \bibnamefont {Nerush}}, \ and\ \bibinfo {author}
  {\bibfnamefont {H.}~\bibnamefont {Ruhl}},\ }\bibfield  {title} {\enquote
  {\bibinfo {title} {Qed cascades induced by circularly polarized laser
  fields},}\ }\href@noop {} {\bibfield  {journal} {\bibinfo  {journal} {Phys.
  Rev. ST Accel. Beams}\ }\textbf {\bibinfo {volume} {14}},\ \bibinfo {pages}
  {054401} (\bibinfo {year} {2011})}\BibitemShut {NoStop}%
\bibitem [{\citenamefont {Ridgers}\ \emph {et~al.}(2014)\citenamefont
  {Ridgers}, \citenamefont {Kirk}, \citenamefont {Duclous}, \citenamefont
  {Blackburn}, \citenamefont {Brady}, \citenamefont {Bennett}, \citenamefont
  {Arber},\ and\ \citenamefont {Bell}}]{Ridgers_2014}%
  \BibitemOpen
  \bibfield  {author} {\bibinfo {author} {\bibfnamefont {C.~P.}\ \bibnamefont
  {Ridgers}}, \bibinfo {author} {\bibfnamefont {J.~G.}\ \bibnamefont {Kirk}},
  \bibinfo {author} {\bibfnamefont {R.}~\bibnamefont {Duclous}}, \bibinfo
  {author} {\bibfnamefont {T.~G.}\ \bibnamefont {Blackburn}}, \bibinfo {author}
  {\bibfnamefont {C.~S.}\ \bibnamefont {Brady}}, \bibinfo {author}
  {\bibfnamefont {K.}~\bibnamefont {Bennett}}, \bibinfo {author} {\bibfnamefont
  {T.~D.}\ \bibnamefont {Arber}}, \ and\ \bibinfo {author} {\bibfnamefont
  {A.~R.}\ \bibnamefont {Bell}},\ }\bibfield  {title} {\enquote {\bibinfo
  {title} {Modelling gamma-ray photon emission and pair production in
  high-intensity laser-matter interactions},}\ }\href@noop {} {\bibfield
  {journal} {\bibinfo  {journal} {J. Compt. Phys.}\ }\textbf {\bibinfo {volume}
  {260}},\ \bibinfo {pages} {273} (\bibinfo {year} {2014})}\BibitemShut
  {NoStop}%
\bibitem [{\citenamefont {Green}\ and\ \citenamefont
  {Harvey}(2015)}]{Green_2015}%
  \BibitemOpen
  \bibfield  {author} {\bibinfo {author} {\bibfnamefont {D.~G.}\ \bibnamefont
  {Green}}\ and\ \bibinfo {author} {\bibfnamefont {C.~N.}\ \bibnamefont
  {Harvey}},\ }\bibfield  {title} {\enquote {\bibinfo {title} {Simla:
  Simulating particle dynamics in intense laser and other electromagnetic
  fields via classical and quantum electrodynamics},}\ }\href@noop {}
  {\bibfield  {journal} {\bibinfo  {journal} {Comp. Phys. Commun.}\ }\textbf
  {\bibinfo {volume} {192}},\ \bibinfo {pages} {313} (\bibinfo {year}
  {2015})}\BibitemShut {NoStop}%
\bibitem [{\citenamefont {Baier}\ \emph {et~al.}(1994)\citenamefont {Baier},
  \citenamefont {Katkov},\ and\ \citenamefont {Strakhovenko}}]{Baier_b_1994}%
  \BibitemOpen
  \bibfield  {author} {\bibinfo {author} {\bibfnamefont {V.~N.}\ \bibnamefont
  {Baier}}, \bibinfo {author} {\bibfnamefont {V.~M.}\ \bibnamefont {Katkov}}, \
  and\ \bibinfo {author} {\bibfnamefont {V.~M.}\ \bibnamefont {Strakhovenko}},\
  }\href@noop {} {\emph {\bibinfo {title} {Electromagnetic Processes at High
  Energies in Oriented Single Crystals}}}\ (\bibinfo  {publisher} {World
  Scientific, Singapore},\ \bibinfo {year} {1994})\BibitemShut {NoStop}%
\bibitem [{\citenamefont {Ritus}(1985)}]{Ritus_1985}%
  \BibitemOpen
  \bibfield  {author} {\bibinfo {author} {\bibfnamefont {V.~I.}\ \bibnamefont
  {Ritus}},\ }\href@noop {} {\bibfield  {journal} {\bibinfo  {journal} {J. Sov.
  Laser Res.}\ }\textbf {\bibinfo {volume} {6}},\ \bibinfo {pages} {497}
  (\bibinfo {year} {1985})}\BibitemShut {NoStop}%
\bibitem [{\citenamefont {Jackson}(1975)}]{Jackson_b_1975}%
  \BibitemOpen
  \bibfield  {author} {\bibinfo {author} {\bibfnamefont {J.~D.}\ \bibnamefont
  {Jackson}},\ }\href@noop {} {\emph {\bibinfo {title} {Classical
  Electrodynamics}}}\ (\bibinfo  {publisher} {Wiley, New York},\ \bibinfo
  {year} {1975})\BibitemShut {NoStop}%
\bibitem [{\citenamefont {Sarachik}\ and\ \citenamefont
  {Schappert}(1970)}]{Sarachik_1970}%
  \BibitemOpen
  \bibfield  {author} {\bibinfo {author} {\bibfnamefont {E.~S.}\ \bibnamefont
  {Sarachik}}\ and\ \bibinfo {author} {\bibfnamefont {G.~T.}\ \bibnamefont
  {Schappert}},\ }\bibfield  {title} {\enquote {\bibinfo {title} {Classical
  theory of the scattering of intense laser radiation by free electrons},}\
  }\href@noop {} {\bibfield  {journal} {\bibinfo  {journal} {Phys. Rev. D}\
  }\textbf {\bibinfo {volume} {1}},\ \bibinfo {pages} {2738--2753} (\bibinfo
  {year} {1970})}\BibitemShut {NoStop}%
\bibitem [{\citenamefont {Bulanov}\ \emph {et~al.}(2010)\citenamefont
  {Bulanov}, \citenamefont {Mur}, \citenamefont {Narozhny}, \citenamefont
  {Nees},\ and\ \citenamefont {Popov}}]{Bulanov_2010_a}%
  \BibitemOpen
  \bibfield  {author} {\bibinfo {author} {\bibfnamefont {S.~S.}\ \bibnamefont
  {Bulanov}}, \bibinfo {author} {\bibfnamefont {V.~D.}\ \bibnamefont {Mur}},
  \bibinfo {author} {\bibfnamefont {N.~B.}\ \bibnamefont {Narozhny}}, \bibinfo
  {author} {\bibfnamefont {J.}~\bibnamefont {Nees}}, \ and\ \bibinfo {author}
  {\bibfnamefont {V.~S.}\ \bibnamefont {Popov}},\ }\bibfield  {title} {\enquote
  {\bibinfo {title} {Multiple colliding electromagnetic pulses: A way to lower
  the threshold of $e^{+}e^{-}$ pair production from vacuum},}\ }\href@noop {}
  {\bibfield  {journal} {\bibinfo  {journal} {Phys. Rev. Lett.}\ }\textbf
  {\bibinfo {volume} {104}},\ \bibinfo {pages} {220404} (\bibinfo {year}
  {2010})}\BibitemShut {NoStop}%
\bibitem [{\citenamefont {Gonoskov}\ \emph {et~al.}(2012)\citenamefont
  {Gonoskov}, \citenamefont {Aiello}, \citenamefont {Heugel},\ and\
  \citenamefont {Leuchs}}]{Gonoskov_2012}%
  \BibitemOpen
  \bibfield  {author} {\bibinfo {author} {\bibfnamefont {Ivan}\ \bibnamefont
  {Gonoskov}}, \bibinfo {author} {\bibfnamefont {Andrea}\ \bibnamefont
  {Aiello}}, \bibinfo {author} {\bibfnamefont {Simon}\ \bibnamefont {Heugel}},
  \ and\ \bibinfo {author} {\bibfnamefont {Gerd}\ \bibnamefont {Leuchs}},\
  }\bibfield  {title} {\enquote {\bibinfo {title} {Dipole pulse theory:
  Maximizing the field amplitude from $4\ensuremath{\pi}$ focused laser
  pulses},}\ }\href {\doibase 10.1103/PhysRevA.86.053836} {\bibfield  {journal}
  {\bibinfo  {journal} {Phys. Rev. A}\ }\textbf {\bibinfo {volume} {86}},\
  \bibinfo {pages} {053836} (\bibinfo {year} {2012})}\BibitemShut {NoStop}%
\bibitem [{\citenamefont {Gonoskov}\ \emph {et~al.}(2017)\citenamefont
  {Gonoskov}, \citenamefont {Bashinov}, \citenamefont {Bastrakov},
  \citenamefont {Efimenko}, \citenamefont {Ilderton}, \citenamefont {Kim},
  \citenamefont {Marklund}, \citenamefont {Meyerov}, \citenamefont {Muraviev},\
  and\ \citenamefont {Sergeev}}]{Gonoskov_2017}%
  \BibitemOpen
  \bibfield  {author} {\bibinfo {author} {\bibfnamefont {A.}~\bibnamefont
  {Gonoskov}}, \bibinfo {author} {\bibfnamefont {A.}~\bibnamefont {Bashinov}},
  \bibinfo {author} {\bibfnamefont {S.}~\bibnamefont {Bastrakov}}, \bibinfo
  {author} {\bibfnamefont {E.}~\bibnamefont {Efimenko}}, \bibinfo {author}
  {\bibfnamefont {A.}~\bibnamefont {Ilderton}}, \bibinfo {author}
  {\bibfnamefont {A.}~\bibnamefont {Kim}}, \bibinfo {author} {\bibfnamefont
  {M.}~\bibnamefont {Marklund}}, \bibinfo {author} {\bibfnamefont
  {I.}~\bibnamefont {Meyerov}}, \bibinfo {author} {\bibfnamefont
  {A.}~\bibnamefont {Muraviev}}, \ and\ \bibinfo {author} {\bibfnamefont
  {A.}~\bibnamefont {Sergeev}},\ }\bibfield  {title} {\enquote {\bibinfo
  {title} {Ultrabright gev photon source via controlled electromagnetic
  cascades in laser-dipole waves},}\ }\href@noop {} {\bibfield  {journal}
  {\bibinfo  {journal} {Phys. Rev. X}\ }\textbf {\bibinfo {volume} {7}},\
  \bibinfo {pages} {041003} (\bibinfo {year} {2017})}\BibitemShut {NoStop}%
\bibitem [{\citenamefont {Magnusson}\ \emph {et~al.}(2019)\citenamefont
  {Magnusson}, \citenamefont {Gonoskov}, \citenamefont {Marklund},
  \citenamefont {Esirkepov}, \citenamefont {Koga}, \citenamefont {Kondo},
  \citenamefont {Kando}, \citenamefont {Bulanov}, \citenamefont {Korn},
  \citenamefont {Geddes}, \citenamefont {Schroeder}, \citenamefont {Esarey},\
  and\ \citenamefont {Bulanov}}]{Magnusson_2019}%
  \BibitemOpen
  \bibfield  {author} {\bibinfo {author} {\bibfnamefont {J.}~\bibnamefont
  {Magnusson}}, \bibinfo {author} {\bibfnamefont {A.}~\bibnamefont {Gonoskov}},
  \bibinfo {author} {\bibfnamefont {M.}~\bibnamefont {Marklund}}, \bibinfo
  {author} {\bibfnamefont {T.~Zh.}\ \bibnamefont {Esirkepov}}, \bibinfo
  {author} {\bibfnamefont {J.~K.}\ \bibnamefont {Koga}}, \bibinfo {author}
  {\bibfnamefont {K.}~\bibnamefont {Kondo}}, \bibinfo {author} {\bibfnamefont
  {M.}~\bibnamefont {Kando}}, \bibinfo {author} {\bibfnamefont {S.~V.}\
  \bibnamefont {Bulanov}}, \bibinfo {author} {\bibfnamefont {G.}~\bibnamefont
  {Korn}}, \bibinfo {author} {\bibfnamefont {C.~G.~R.}\ \bibnamefont {Geddes}},
  \bibinfo {author} {\bibfnamefont {C.~B.}\ \bibnamefont {Schroeder}}, \bibinfo
  {author} {\bibfnamefont {E.}~\bibnamefont {Esarey}}, \ and\ \bibinfo {author}
  {\bibfnamefont {S.~S.}\ \bibnamefont {Bulanov}},\ }\bibfield  {title}
  {\enquote {\bibinfo {title} {Multiple colliding laser pulses as a basis for
  studying high-field high-energy physics},}\ }\href {\doibase
  10.1103/PhysRevA.100.063404} {\bibfield  {journal} {\bibinfo  {journal}
  {Phys. Rev. A}\ }\textbf {\bibinfo {volume} {100}},\ \bibinfo {pages}
  {063404} (\bibinfo {year} {2019})}\BibitemShut {NoStop}%
\bibitem [{\citenamefont {Gonoskov}\ \emph {et~al.}(2014)\citenamefont
  {Gonoskov}, \citenamefont {Bashinov}, \citenamefont {Gonoskov}, \citenamefont
  {Harvey}, \citenamefont {Ilderton}, \citenamefont {Kim}, \citenamefont
  {Marklund}, \citenamefont {Mourou},\ and\ \citenamefont
  {Sergeev}}]{Gonoskov_2014}%
  \BibitemOpen
  \bibfield  {author} {\bibinfo {author} {\bibfnamefont {A.}~\bibnamefont
  {Gonoskov}}, \bibinfo {author} {\bibfnamefont {A.}~\bibnamefont {Bashinov}},
  \bibinfo {author} {\bibfnamefont {I.}~\bibnamefont {Gonoskov}}, \bibinfo
  {author} {\bibfnamefont {C.}~\bibnamefont {Harvey}}, \bibinfo {author}
  {\bibfnamefont {A.}~\bibnamefont {Ilderton}}, \bibinfo {author}
  {\bibfnamefont {A.}~\bibnamefont {Kim}}, \bibinfo {author} {\bibfnamefont
  {M.}~\bibnamefont {Marklund}}, \bibinfo {author} {\bibfnamefont
  {G.}~\bibnamefont {Mourou}}, \ and\ \bibinfo {author} {\bibfnamefont
  {A.}~\bibnamefont {Sergeev}},\ }\bibfield  {title} {\enquote {\bibinfo
  {title} {Anomalous radiative trapping in laser fields of extreme
  intensity},}\ }\href@noop {} {\bibfield  {journal} {\bibinfo  {journal}
  {Phys. Rev. Lett.}\ }\textbf {\bibinfo {volume} {113}},\ \bibinfo {pages}
  {014801} (\bibinfo {year} {2014})}\BibitemShut {NoStop}%
\bibitem [{\citenamefont {Kirk}(2016)}]{Kirk_2016}%
  \BibitemOpen
  \bibfield  {author} {\bibinfo {author} {\bibfnamefont {J~G}\ \bibnamefont
  {Kirk}},\ }\bibfield  {title} {\enquote {\bibinfo {title} {Radiative trapping
  in intense laser beams},}\ }\href@noop {} {\bibfield  {journal} {\bibinfo
  {journal} {Plasma Phys. Cont. Fus.}\ }\textbf {\bibinfo {volume} {58}},\
  \bibinfo {pages} {085005} (\bibinfo {year} {2016})}\BibitemShut {NoStop}%
\bibitem [{\citenamefont {Kirk}\ \emph {et~al.}(2009)\citenamefont {Kirk},
  \citenamefont {Bell},\ and\ \citenamefont {Arka}}]{Kirk_2009}%
  \BibitemOpen
  \bibfield  {author} {\bibinfo {author} {\bibfnamefont {J~G}\ \bibnamefont
  {Kirk}}, \bibinfo {author} {\bibfnamefont {A~R}\ \bibnamefont {Bell}}, \ and\
  \bibinfo {author} {\bibfnamefont {I}~\bibnamefont {Arka}},\ }\bibfield
  {title} {\enquote {\bibinfo {title} {Pair production in counter-propagating
  laser beams},}\ }\href@noop {} {\bibfield  {journal} {\bibinfo  {journal}
  {Plasma Phys. Contr. F.}\ }\textbf {\bibinfo {volume} {51}},\ \bibinfo
  {pages} {085008} (\bibinfo {year} {2009})}\BibitemShut {NoStop}%
\bibitem [{\citenamefont {Grismayer}\ \emph {et~al.}(2016)\citenamefont
  {Grismayer}, \citenamefont {Vranic}, \citenamefont {Martins}, \citenamefont
  {Fonseca},\ and\ \citenamefont {Silva}}]{Grismayer_2016}%
  \BibitemOpen
  \bibfield  {author} {\bibinfo {author} {\bibfnamefont {T.}~\bibnamefont
  {Grismayer}}, \bibinfo {author} {\bibfnamefont {M.}~\bibnamefont {Vranic}},
  \bibinfo {author} {\bibfnamefont {J.~L.}\ \bibnamefont {Martins}}, \bibinfo
  {author} {\bibfnamefont {R.~A.}\ \bibnamefont {Fonseca}}, \ and\ \bibinfo
  {author} {\bibfnamefont {L.~O.}\ \bibnamefont {Silva}},\ }\bibfield  {title}
  {\enquote {\bibinfo {title} {Laser absorption via quantum electrodynamics
  cascades in counter propagating laser pulses},}\ }\href@noop {} {\bibfield
  {journal} {\bibinfo  {journal} {Phys. Plasmas}\ }\textbf {\bibinfo {volume}
  {23}},\ \bibinfo {pages} {056706} (\bibinfo {year} {2016})}\BibitemShut
  {NoStop}%
\bibitem [{\citenamefont {Jirka}\ \emph {et~al.}(2016)\citenamefont {Jirka},
  \citenamefont {Klimo}, \citenamefont {Bulanov}, \citenamefont {Esirkepov},
  \citenamefont {Gelfer}, \citenamefont {Bulanov}, \citenamefont {Weber},\ and\
  \citenamefont {Korn}}]{Jirka_2016}%
  \BibitemOpen
  \bibfield  {author} {\bibinfo {author} {\bibfnamefont {M.}~\bibnamefont
  {Jirka}}, \bibinfo {author} {\bibfnamefont {O.}~\bibnamefont {Klimo}},
  \bibinfo {author} {\bibfnamefont {S.~V.}\ \bibnamefont {Bulanov}}, \bibinfo
  {author} {\bibfnamefont {T.~Zh.}\ \bibnamefont {Esirkepov}}, \bibinfo
  {author} {\bibfnamefont {E.}~\bibnamefont {Gelfer}}, \bibinfo {author}
  {\bibfnamefont {S.~S.}\ \bibnamefont {Bulanov}}, \bibinfo {author}
  {\bibfnamefont {S.}~\bibnamefont {Weber}}, \ and\ \bibinfo {author}
  {\bibfnamefont {G.}~\bibnamefont {Korn}},\ }\bibfield  {title} {\enquote
  {\bibinfo {title} {Electron dynamics and $\ensuremath{\gamma}$ and
  ${e}^{\ensuremath{-}}{e}^{+}$ production by colliding laser pulses},}\
  }\href@noop {} {\bibfield  {journal} {\bibinfo  {journal} {Phys. Rev. E}\
  }\textbf {\bibinfo {volume} {93}},\ \bibinfo {pages} {023207} (\bibinfo
  {year} {2016})}\BibitemShut {NoStop}%
\bibitem [{\citenamefont {Gong}\ \emph {et~al.}(2017)\citenamefont {Gong},
  \citenamefont {Hu}, \citenamefont {Shou}, \citenamefont {Qiao}, \citenamefont
  {Chen}, \citenamefont {He}, \citenamefont {Bulanov}, \citenamefont
  {Esirkepov}, \citenamefont {Bulanov},\ and\ \citenamefont {Yan}}]{Gong_2017}%
  \BibitemOpen
  \bibfield  {author} {\bibinfo {author} {\bibfnamefont {Z.}~\bibnamefont
  {Gong}}, \bibinfo {author} {\bibfnamefont {R.~H.}\ \bibnamefont {Hu}},
  \bibinfo {author} {\bibfnamefont {Y.~R.}\ \bibnamefont {Shou}}, \bibinfo
  {author} {\bibfnamefont {B.}~\bibnamefont {Qiao}}, \bibinfo {author}
  {\bibfnamefont {C.~E.}\ \bibnamefont {Chen}}, \bibinfo {author}
  {\bibfnamefont {X.~T.}\ \bibnamefont {He}}, \bibinfo {author} {\bibfnamefont
  {S.~S.}\ \bibnamefont {Bulanov}}, \bibinfo {author} {\bibfnamefont {T.~Zh.}\
  \bibnamefont {Esirkepov}}, \bibinfo {author} {\bibfnamefont {S.~V.}\
  \bibnamefont {Bulanov}}, \ and\ \bibinfo {author} {\bibfnamefont {X.~Q.}\
  \bibnamefont {Yan}},\ }\bibfield  {title} {\enquote {\bibinfo {title}
  {High-efficiency $\ensuremath{\gamma}$-ray flash generation via
  multiple-laser scattering in ponderomotive potential well},}\ }\href@noop {}
  {\bibfield  {journal} {\bibinfo  {journal} {Phys. Rev. E}\ }\textbf {\bibinfo
  {volume} {95}},\ \bibinfo {pages} {013210} (\bibinfo {year}
  {2017})}\BibitemShut {NoStop}%
\bibitem [{\citenamefont {Grismayer}\ \emph {et~al.}(2017)\citenamefont
  {Grismayer}, \citenamefont {Vranic}, \citenamefont {Martins}, \citenamefont
  {Fonseca},\ and\ \citenamefont {Silva}}]{Grismayer_2017}%
  \BibitemOpen
  \bibfield  {author} {\bibinfo {author} {\bibfnamefont {T.}~\bibnamefont
  {Grismayer}}, \bibinfo {author} {\bibfnamefont {M.}~\bibnamefont {Vranic}},
  \bibinfo {author} {\bibfnamefont {J.~L.}\ \bibnamefont {Martins}}, \bibinfo
  {author} {\bibfnamefont {R.~A.}\ \bibnamefont {Fonseca}}, \ and\ \bibinfo
  {author} {\bibfnamefont {L.~O.}\ \bibnamefont {Silva}},\ }\bibfield  {title}
  {\enquote {\bibinfo {title} {Seeded qed cascades in counterpropagating laser
  pulses},}\ }\href {\doibase 10.1103/PhysRevE.95.023210} {\bibfield  {journal}
  {\bibinfo  {journal} {Phys. Rev. E}\ }\textbf {\bibinfo {volume} {95}},\
  \bibinfo {pages} {023210} (\bibinfo {year} {2017})}\BibitemShut {NoStop}%
\bibitem [{\citenamefont {Dellweg}\ and\ \citenamefont
  {M\"uller}(2017)}]{Mueller_2017}%
  \BibitemOpen
  \bibfield  {author} {\bibinfo {author} {\bibfnamefont {M.M.}\ \bibnamefont
  {Dellweg}}\ and\ \bibinfo {author} {\bibfnamefont {C.}~\bibnamefont
  {M\"uller}},\ }\bibfield  {title} {\enquote {\bibinfo {title}
  {Spin-polarizing interferometric beam splitter for free electrons},}\
  }\href@noop {} {\bibfield  {journal} {\bibinfo  {journal} {Phys. Rev. Lett.}\
  }\textbf {\bibinfo {volume} {118}},\ \bibinfo {pages} {070403} (\bibinfo
  {year} {2017})}\BibitemShut {NoStop}%
\bibitem [{\citenamefont {Brady}\ \emph {et~al.}(2012)\citenamefont {Brady},
  \citenamefont {Ridgers}, \citenamefont {Arber}, \citenamefont {Bell},\ and\
  \citenamefont {Kirk}}]{Brady_2012}%
  \BibitemOpen
  \bibfield  {author} {\bibinfo {author} {\bibfnamefont {C.~S.}\ \bibnamefont
  {Brady}}, \bibinfo {author} {\bibfnamefont {C.~P.}\ \bibnamefont {Ridgers}},
  \bibinfo {author} {\bibfnamefont {T.~D.}\ \bibnamefont {Arber}}, \bibinfo
  {author} {\bibfnamefont {A.~R.}\ \bibnamefont {Bell}}, \ and\ \bibinfo
  {author} {\bibfnamefont {J.~G.}\ \bibnamefont {Kirk}},\ }\bibfield  {title}
  {\enquote {\bibinfo {title} {Laser absorption in relativistically underdense
  plasmas by synchrotron radiation},}\ }\href@noop {} {\bibfield  {journal}
  {\bibinfo  {journal} {Phys. Rev. Lett.}\ }\textbf {\bibinfo {volume} {109}},\
  \bibinfo {pages} {245006} (\bibinfo {year} {2012})}\BibitemShut {NoStop}%
\bibitem [{\citenamefont {{Di Piazza}}\ \emph {et~al.}(2018)\citenamefont {{Di
  Piazza}}, \citenamefont {Tamburini}, \citenamefont {Meuren},\ and\
  \citenamefont {Keitel}}]{DiPiazza_2018}%
  \BibitemOpen
  \bibfield  {author} {\bibinfo {author} {\bibfnamefont {A}~\bibnamefont {{Di
  Piazza}}}, \bibinfo {author} {\bibfnamefont {M}~\bibnamefont {Tamburini}},
  \bibinfo {author} {\bibfnamefont {S}~\bibnamefont {Meuren}}, \ and\ \bibinfo
  {author} {\bibfnamefont {C~H}\ \bibnamefont {Keitel}},\ }\bibfield  {title}
  {\enquote {\bibinfo {title} {{Implementing nonlinear Compton scattering
  beyond the local-constant-field approximation}},}\ }\href@noop {} {\bibfield
  {journal} {\bibinfo  {journal} {Phys. Rev. A}\ }\textbf {\bibinfo {volume}
  {98}},\ \bibinfo {pages} {012134} (\bibinfo {year} {2018})}\BibitemShut
  {NoStop}%
\bibitem [{\citenamefont {{Di Piazza}}\ \emph {et~al.}(2019)\citenamefont {{Di
  Piazza}}, \citenamefont {Tamburini}, \citenamefont {Meuren},\ and\
  \citenamefont {Keitel}}]{DiPiazza_2019}%
  \BibitemOpen
  \bibfield  {author} {\bibinfo {author} {\bibfnamefont {A}~\bibnamefont {{Di
  Piazza}}}, \bibinfo {author} {\bibfnamefont {M}~\bibnamefont {Tamburini}},
  \bibinfo {author} {\bibfnamefont {S}~\bibnamefont {Meuren}}, \ and\ \bibinfo
  {author} {\bibfnamefont {C~H}\ \bibnamefont {Keitel}},\ }\bibfield  {title}
  {\enquote {\bibinfo {title} {{Improved local-constant-field approximation for
  strong-field QED codes}},}\ }\href@noop {} {\bibfield  {journal} {\bibinfo
  {journal} {Phys. Rev. A}\ }\textbf {\bibinfo {volume} {99}},\ \bibinfo
  {pages} {022125} (\bibinfo {year} {2019})}\BibitemShut {NoStop}%
\bibitem [{\citenamefont {Ilderton}\ \emph {et~al.}(2019)\citenamefont
  {Ilderton}, \citenamefont {King},\ and\ \citenamefont
  {Seipt}}]{Ilderton_2019}%
  \BibitemOpen
  \bibfield  {author} {\bibinfo {author} {\bibfnamefont {A}~\bibnamefont
  {Ilderton}}, \bibinfo {author} {\bibfnamefont {B}~\bibnamefont {King}}, \
  and\ \bibinfo {author} {\bibfnamefont {D}~\bibnamefont {Seipt}},\ }\bibfield
  {title} {\enquote {\bibinfo {title} {{Extended locally constant field
  approximation for nonlinear Compton scattering}},}\ }\href@noop {} {\bibfield
   {journal} {\bibinfo  {journal} {Phys. Rev. A}\ }\textbf {\bibinfo {volume}
  {99}},\ \bibinfo {pages} {042121} (\bibinfo {year} {2019})}\BibitemShut
  {NoStop}%
\bibitem [{\citenamefont {Baier}\ and\ \citenamefont
  {Katkov}(1968)}]{Katkov_1968}%
  \BibitemOpen
  \bibfield  {author} {\bibinfo {author} {\bibfnamefont {V.~N.}\ \bibnamefont
  {Baier}}\ and\ \bibinfo {author} {\bibfnamefont {V.~M.}\ \bibnamefont
  {Katkov}},\ }\href@noop {} {\bibfield  {journal} {\bibinfo  {journal} {Sov.
  Phys. JETP}\ }\textbf {\bibinfo {volume} {26}},\ \bibinfo {pages} {854}
  (\bibinfo {year} {1968})}\BibitemShut {NoStop}%
\bibitem [{Sup()}]{Supp}%
  \BibitemOpen
  \href@noop {} {}\bibinfo {howpublished} {See the Supplemental Materials for
  the details of calculations of radiation spectra.}\BibitemShut {Stop}%
\end{thebibliography}%

\clearpage


\section*{Supplementary material}

# Supplemental Materials to the paper
*"New class of violation of local constant field approximation in intense crossed laser pulse scenarios"*

## I. RADIATION FORMATION TIME

In the paper, we employ the semiclassical Baier-Katkov approach [1] to calculate the radiation spectra in a strong external field. The phase of a photon emission is a crucial parameter in the formalism, determining the interference of radiation emerging from different points of the trajectory. It is given by

$$\psi = \frac{\varepsilon}{\varepsilon'} k \cdot x(t),$$ (1)

where $k_\mu = \omega(1, \boldsymbol{n})$ is the emitted photon four-wavevector. Introducing the definition of $u = \omega/(\varepsilon - \omega)$, we have

$$\psi = m\gamma u \left[ t - \boldsymbol{n} \cdot \boldsymbol{x}(t) \right].$$ (2)

The trajectory in the vicinity of $t_0$ can be represented as

$$\boldsymbol{x}(t_0 + \tau) = \boldsymbol{x}_0 + \boldsymbol{v}(t_0)\tau + \int_0^\tau d\tau' \left[ \boldsymbol{v}(t_0 + \tau') - \boldsymbol{v}(t_0) \right],$$ (3)

with $\boldsymbol{x}_0 \equiv \boldsymbol{x}(t_0)$. Therefore, the phase reads

$$
\begin{aligned}
\psi &= mu\gamma \left[ (1 - v(t_0))\,\tau - \boldsymbol{v}(t_0) \cdot \int_0^\tau d\tau' \left[ \boldsymbol{v}(t_0 + \tau') - \boldsymbol{v}(t_0) \right] \right] \\
&\approx mu\gamma \left[ \frac{1}{2\gamma^2}\tau + \int_0^\tau d\tau' \Lambda(t_0, \tau') \right],
\end{aligned}
$$ (4)

where $\boldsymbol{n} \approx \boldsymbol{v}(t_0)$ is assumed, i.e. forward emission for the ultra-relativistic electron, and $\Lambda(t_0, \tau) \equiv -\boldsymbol{v}(t_0) \cdot \left[ \boldsymbol{v}(t_0 + \tau) - \boldsymbol{v}(t_0) \right]$. The constant phase $m\gamma u(t_0 - \boldsymbol{n} \cdot \boldsymbol{x}_0)$ is omitted as it does not affect the interference. In order to understand the physical meaning of $\Lambda(t_0, \tau)$, we rewrite it in the following way

$$\Lambda(t_0, \tau) = -v^2(t_0) \left( \cos\theta(\tau) - 1 \right) \approx \frac{1}{2} v^2(t_0)\theta^2(\tau),$$ (5)

where $\theta(\tau) \ll 1$ is the angle between $\boldsymbol{v}(t_0 + \tau)$ and $\boldsymbol{v}(t_0)$. Having expressed the phase $\psi$ in terms of $\theta$ and recalling the well-known fact that the radiation formation interval corresponds to $\theta(\tau) \lesssim 1/\gamma$, the formation time may be instantly obtained.

Let us find an explicit expression for the formation time according to the LCFA. This approximation corresponds to expanding the velocity up to the $\tau^2$ order, namely

$$\Lambda(t_0, \tau) = -\frac{1}{2} \left[ \boldsymbol{v}(t_0) \cdot \ddot{\boldsymbol{v}}(t_0) \right] \tau^2 = \frac{1}{2} |\dot{v}|^2 \tau^2 = \frac{m^2 \chi^2}{2\gamma^4} \tau^2,$$ (6)

where we have assumed that the force acting on the electron is transverse $\boldsymbol{v}(t_0) \cdot \dot{\boldsymbol{v}}(t_0) = 0$, and the definition of the quantum parameter $\chi = \gamma^2 |\dot{v}| m$ was employed. Combining Eq. (5) and (6), the time dependence of the angle according to the LCFA reads

$$\theta_L(\tau) = \frac{m\chi}{\gamma^2} \tau.$$ (7)

Therefore, the LCFA formation time, defined by $\theta(t_f^L) = 2/\gamma$, is given by

$$t_f^L = \frac{2\gamma}{m\chi}.$$ (8)

Accordingly, in order to determine for a given trajectory, whether a coincidence with LCFA is to be expected or not, one should examine the temporal behavior of the angle $\theta(\tau)$ during the emission interval. A linear dependence indicates a good agreement with LCFA.

## II. CPW SCENARIO

### A. Radiation spectral distribution

Using the trajectory given in the main text, the Baier-Katkov formula for the radiation spectral distribution can be represented in the following form:

$$
\begin{aligned}
\frac{dI}{d\omega d\varphi} = \frac{\alpha\omega}{4\pi \bar{v}_z(1 + u)} \sum_{s_r} \sum_{s_l} \\
\left[ -\left( 2 + 2u + u^2 \right) |\mathcal{M}_\mu|^2 + \left( \frac{u}{\gamma} \right)^2 |\mathcal{M}_0|^2 \right].
\end{aligned}
$$ (9)

In the following, the matrix element $\mathcal{M}_\mu$ is given as a function of $s_l, s_r, u, \varphi$. We start by introducing the quantities

$$
\begin{aligned}
B_0(s, z, \varphi) &= J_s(z) e^{-is\varphi}, \\
B_1(s, z, \varphi) &= \left[ \frac{s}{z} J_s(z) \cos\varphi + i J_s'(z) \sin\varphi \right] e^{-is\varphi}, \\
B_2(s, z, \varphi) &= \left[ \frac{s}{z} J_s(z) \sin\varphi - i J_s'(z) \cos\varphi \right] e^{-is\varphi},
\end{aligned}
$$ (10)

where $J_s(z), J_s'(z)$ are the Bessel function and its first derivative, respectively. In terms of these functions, the various components of the matrix element take the form

$$
\begin{aligned}
\mathcal{M}_t &= \sum_{s_3} B_0(\boldsymbol{1}) B_0(\boldsymbol{2}) B_0(\boldsymbol{3}), \\
\mathcal{M}_x &= \frac{m}{\varepsilon} \sum_{s_3} \left[ \xi_1 B_0(\boldsymbol{2}) B_1(\boldsymbol{1}) + \xi_2 B_0(\boldsymbol{1}) B_1(\boldsymbol{2}) \right] B_0(\boldsymbol{3}), \\
\mathcal{M}_y &= \frac{m}{\varepsilon} \sum_{s_3} m \left[ \xi_1 B_0(\boldsymbol{2}) B_2(\boldsymbol{1}) + \xi_2 B_0(\boldsymbol{1}) B_2(\boldsymbol{2}) \right] B_0(\boldsymbol{3}), \\
\mathcal{M}_z &= \sum_{s_3} B_0(\boldsymbol{1}) B_0(\boldsymbol{2}) \left[ \bar{v}_z B_0(\boldsymbol{3}) - \frac{m^2 \xi_1 \xi_2}{\bar{v}_z \varepsilon^2} B_1(\boldsymbol{3}) \right],
\end{aligned}
$$ (11)

where $\boldsymbol{1} \equiv (s_1, z_1, \varphi)$, $\boldsymbol{2} \equiv (s_2, z_2, \varphi)$ and $\boldsymbol{3} \equiv (s_3, z_3, 0)$. The indices $s_1, s_2, s_3$ are related to $s_r, s_l$ that appear in the final emission expression by

$$s_1 \equiv s_l - s_3 \quad, s_2 \equiv s_r + s_3.$$ (12)

The arguments of the Bessel function read

$$z_1 \equiv \frac{m\xi_1 u \sin\vartheta}{\omega_1}, z_2 \equiv \frac{m\xi_2 u \sin\vartheta}{\omega_2}, z_3 \equiv \frac{m^2\xi_1\xi_2 u}{\bar{v}_z \Delta\omega\varepsilon}\cos\vartheta .$$

The angle $\vartheta$ is expressed in terms of $u, s_l, s_r$ according to

$$\cos\vartheta = \frac{1}{\bar{v}_z}\left[1 - \frac{1}{\varepsilon u}(s_l\omega_1 + s_r\omega_2)\right]. \quad (13)$$

### B. Analytical expressions for $\chi$ and $t_c$

The acceleration, corresponding to the trajectory given in the paper, reads

$$\dot{v}_x = -\frac{m\xi_1\omega_1}{\varepsilon}\sin\omega_1 t - \frac{m\xi_2\omega_2}{\varepsilon}\sin\omega_2 t ,$$

$$\dot{v}_y = \frac{m\xi_1\omega_1}{\varepsilon}\cos\omega_1 t + \frac{m\xi_2\omega_2}{\varepsilon}\cos\omega_2 t , \quad (14)$$

$$\dot{v}_z = \frac{2\omega m^2\xi_1\xi_2}{\varepsilon^2}\sin\Delta\omega t .$$

Hence

$$\chi^2 = \frac{\gamma^4\dot{v}^2}{m^2} = \chi_1^2 + \chi_2^2 - 2\chi_1\chi_2\cos\Delta\omega t + \frac{\xi_1^2\chi_2^2}{\gamma^2}\sin^2\Delta\omega t. \quad (15)$$

The last term is negligible, since $\xi_1 \ll \gamma$. Thus, the expression appearing in the paper is obtained.

$$\chi \approx \sqrt{\chi_1^2 + \chi_2^2 - 2\chi_1\chi_2\cos\Delta\omega t}. \quad (16)$$

We further calculate the characteristic time of the electron trajectory $t_c = |\mathbf{F}|/|\dot{\mathbf{F}}|$, where $\mathbf{F}$ is the Lorentz force. Since the energy is constant, this quantity may be equivalently written as $t_c = |\dot{v}|/|\ddot{v}|$. The acceleration $\dot{v}$ was already found, and the denominator $\ddot{v}$ is straightforwardly obtained by taking a derivative of Eq. (14).

$$|\ddot{v}| \approx \frac{m}{\gamma^2}\sqrt{\omega_1^2\chi_1^2 + \omega_2^2\chi_2^2 - 2\chi_1\chi_2\omega_1\omega_2\sin\Delta\omega t}, \quad (17)$$

and hence the trajectory characteristic time takes the form

$$t_c \approx \frac{\chi(t)}{\sqrt{\omega_1^2\chi_1^2 + \omega_2^2\chi_2^2 - 2\chi_1\chi_2\omega_1\omega_2\sin\Delta\omega t}}. \quad (18)$$

In the cases $\chi_1 \sim \chi_2$ and $\omega_2 \gg \omega_1$, the denominator can be approximated by $\omega_2\chi_2$, leading to

$$t_c \approx \frac{\chi(t)}{\omega_2\chi_2}. \quad (19)$$

### C. Additional spectra for the $\xi_1 = 100, \xi_2 = 2$ case

In the paper, we concentrated on the most interesting case of $\chi_1 \approx \chi_2$, where the emission is influenced by both beams. Here

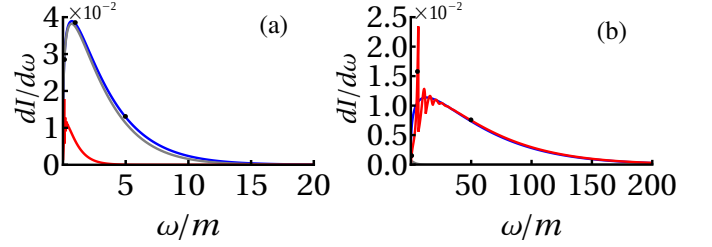

Figure 1. Emission spectrum for the CPW scenario with $\xi_1 = 100, \xi_2 = 2$. (a) corresponds to $\varepsilon = 2m_*$ and (b) to $\varepsilon = 16m_*$. The color scheme is the same as in Fig. 2 of the paper.

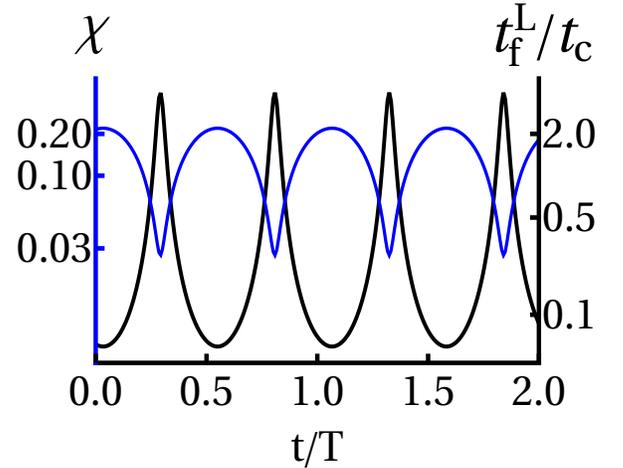

Figure 2. The ratio $t_f^L/t_c$ and the quantum parameter $\chi$ as a function of time for the case of $\xi_1 = 500$ and $\xi_2 = 10$ presented in the paper.

we show that in the limits when $\chi_1 \gg \chi_2$ or vice versa, one beam dominates and, accordingly, the full spectrum follows the corresponding single laser field results. The spectrum calculated for $\varepsilon = 2m_*$ is shown in Fig. 1(a). For the given parameters $\chi_1 = 8.13 \times 10^{-2}$ is substantially larger than $\chi_2 = 2.26 \times 10^{-2}$ and, therefore, the emission spectrum (black) closely follows the one corresponding to the first beam (gray). Since $\xi_1 = 100$ this spectrum also coincides with the LCFA one (blue). The numerical calculation (black dots) agrees with the analytical one (black).

For the high energy case ($\varepsilon = 16m_*$), one finds $\chi_1 = 9.49 \times 10^{-4}$ and $\chi_2 = 1.94 \times 10^{-2}$ and thus the spectrum is similar to the second beam emission, see Fig. 1(b). The numerically calculated emission (black dots) is identical to the emission corresponding to a single beam with $\xi_2$ (red curve). The harmonic structure of the red curve is well reproduced by the total spectrum (black dots). The LCFA result (blue curve) follows the harmonic-averaged emission, as expected.

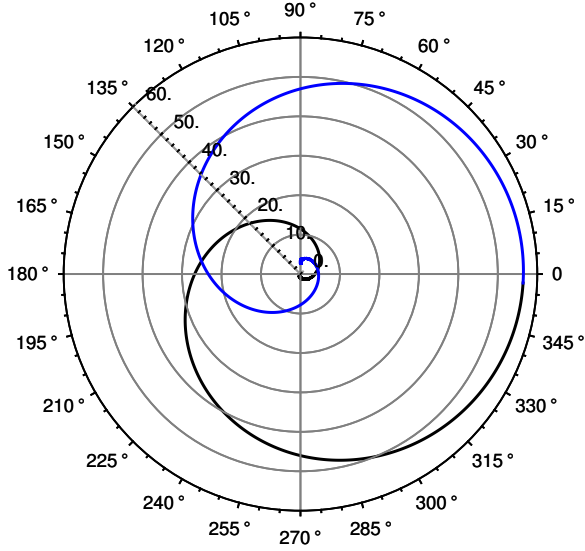

Figure 3. The trajectory of the particle in the angular plane ($\vartheta - \varphi$). The blue (black) curve designates the trajectory during the rise (fall) of the pulse. The spectrum considered in the paper is for $\varphi = 3\pi/4$.

### D. $t_f^L/t_c$ for the $\xi_1 = 500, \xi_2 = 10$ case

Fig. 3(c,d) of the paper demonstrates that even for $\xi_1 = 500$ and $\xi_2 = 10$ the total spectrum does not agree with the LCFA calculation. In order to better understand the underlying reason of this deviation, the quantity $t_f^L/t_c$ is presented in Fig. 2. As in the $\xi_1 = 100, \xi_2 = 2$ case, depicted in Fig. 3(a) of the paper, the ratio $t_f^L/t_c$ is higher than 1 in the trajectory part corresponding to low $\chi$ value. Thus, a discrepancy with respect to the LCFA is obviously expected. However, the relative weight of this trajectory part is smaller as compared to the above mentioned case, and therefore so is the total deviation from the LCFA spectrum (as shown in Fig. 4(d) of the paper).

## III. ULTRASHORT PULSE SCENARIO

### A. Pulse shape

In order to specify the spatial and temporal shape for the ultra-short pulse used in the paper, the following quantities are introduced

$$X \equiv \frac{x}{w_0}, Y \equiv \frac{y}{w_0}, Z \equiv \frac{z}{z_r}, z_r \equiv \frac{\omega w_0^2}{2}, \quad (20)$$

and

$$f \equiv \frac{i}{i + Z}, \rho \equiv \sqrt{X^2 + Y^2}. \quad (21)$$

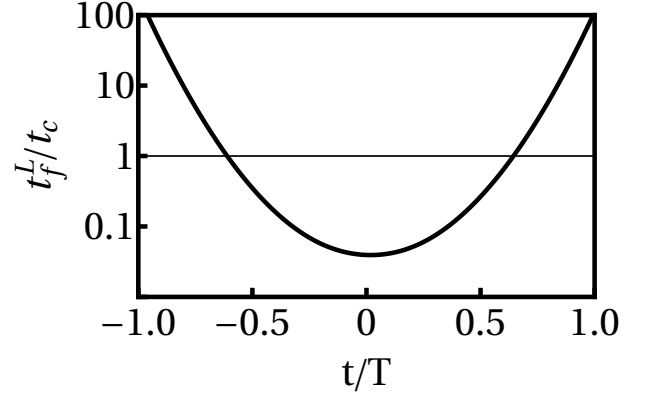

Figure 4. The ratio $t_f^L/t_c$ as a function of time for the ultra-short laser pulse scenario.

The vector potential corresponding to this ultra-short and tightly-focused pulse reads [2]

$$A_x = A_0 g(\eta) f e^{-f\rho^2} e^{i\eta} \times$$
$$\left(1 + \epsilon^2 \left[\frac{f}{2} - \frac{f^3 \rho^4}{4}\right] + \epsilon^4 \left[\frac{3f^2}{8} - \frac{3f^4 \rho^4}{16} - \frac{f^5 \rho^6}{8} + \frac{f^6 \rho^8}{32}\right]\right) \quad (22)$$

where $A_0$ denotes the amplitude and $\eta \equiv k \cdot x$ and $g(\eta) = e^{-\frac{\eta^2}{2\sigma^2}}$ is the temporal envelope. Since we consider a circular polarization, the $y$ component is given by $A_y = iA_x$. The electromagnetic fields can thus be derived from the above vector potential by

$$\mathbf{E} = -i\omega \mathbf{A} - \frac{i}{\omega} \mathbf{\nabla}(\mathbf{\nabla} \cdot \mathbf{A}), \quad (23)$$
$$\mathbf{B} = \mathbf{\nabla} \times \mathbf{A}.$$

### B. The particle trajectory in the $\vartheta - \varphi$ plane

In the paper, we argue that the new mechanism of deviating from the LCFA prediction can also occur in the ultrashort pulse scenario because of the rapid rising and falling tails of the pulse. In the following a further intuitive explanation is provided. Fig. 3 illustrates the trajectory of an electron in the $\vartheta - \varphi$ plane corresponding to the parameters used in the paper for Fig. 5. The polar axis stands for the angle $\varphi = 3\pi/4$, for which the angular resolved spectrum in the paper is calculated. One may see that the trajectory crosses this line three times, which implies that the angular resolved spectrum should feature three peaks, as presented in the paper (see Fig. 5(b) of the paper). In the paper it was demonstrated that the region corresponding to the high $\vartheta$ values are well described by the LCFA, and the deviation stems from the small $\vartheta$ direction. Looking closely at the trajectory in Fig. 3 one may indeed see the dramatic change in the trajectory at this point, in agreement with the $\theta(t)$ curve appearing in Fig. 5(c) of the paper.

### C. $t_f^L/t_c$ for the ultra-short laser pulse case

The deviation from the LCFA can also be proved from the ratio $t_f^L/t_c$ depicted in Fig. 4. From the figure, one can see that the ratio is much large than unity at the beginning and end of the pulse. This indicates the failure of the LCFA prediction. As we argued in the paper, the surprising point is that the failure can induce a deviation in the high energy domain, even though $\chi$ is much smaller at the beginning and end of the pulse.

[1] V. N. Baier, V. M. Katkov, and V. M. Strakhovenko, *Electromagnetic Processes at High Energies in Oriented Single Crystals* (World Scientific, Singapore, 1994).

[2] Yousef I. Salamin, Guido R. Mocken, and Christoph H. Keitel, "Electron scattering and acceleration by a tightly focused laser beam," Phys. Rev. ST Accel. Beams **5**, 101301 (2002).



\end{document}